\documentclass[twocolumn]{aastex62}

\usepackage{xcolor}

\definecolor{darkbrown}{HTML}{b15a00}
\definecolor{darkblue}{HTML}{001166}

\newcommand{\y}{\checkmark}
\newcommand{\n}{$\times$}

\newcommand{\treinf}{t_{\rm reinf}}
\newcommand{\tstop}{t_{\rm stop}}

\renewcommand{\vr}{v_{\rm r}}
\newcommand{\vk}{v_{\rm k}}
\newcommand{\cs}{c_{\rm s}}
\newcommand{\rhos}{\rho_{\rm s}}
\newcommand{\rhop}{\rho_{\rm p}}

\newcommand{\Omegainv}{\Omega^{-1}}

\newcommand{\Sigmap}{\Sigma_{\rm p}}

\newcommand{\bmath}[1]{\mbox{\boldmath{$#1$}}}

\newcommand{\del}{{\bf \nabla}}

\graphicspath{{./}{plots/}}

\submitjournal{ApJ}

\shorttitle{Planetesimal Formation in Disk Rings}
\shortauthors{Carrera et al.}

\begin{document}

\title{Protoplanetary Disk Rings as Sites for Planetesimal Formation}

\correspondingauthor{Daniel Carrera}
\email{dcarrera@gmail.com}

\author[0000-0001-6259-3575]{Daniel Carrera}
\affiliation{Department of Physics and Astronomy, Iowa State University, Ames, IA, 50010, USA}

\author[0000-0002-3771-8054]{Jacob B. Simon}
\affiliation{Department of Physics and Astronomy, Iowa State University, Ames, IA 50010, USA}
\affiliation{JILA, University of Colorado and NIST, 440 UCB, Boulder, CO 80309, USA}
\affiliation{Department of Space Studies, Southwest Research Institute, Boulder, CO 80302, USA}

\author[0000-0001-9222-4367]{Rixin Li}
\affiliation{Department of Astronomy and Steward Observatory, University of Arizona, 933 North Cherry Avenue, Tucson, AZ 85721, USA}

\author{Katherine A. Kretke}
\affiliation{Department of Space Studies, Southwest Research Institute, Boulder, CO 80302, USA}

\author[0000-0002-8227-5467]{Hubert Klahr}
\affiliation{Max-Planck-Institut f{\"u}r Astronomie: Heidelberg, Baden-W{\"u}rttemberg}

\begin{abstract}
Axisymmetric dust rings are a ubiquitous feature of young protoplanetary disks. These rings are likely caused by pressure bumps in the gas profile; a small bump can induce a traffic jam-like pattern in the dust density, while a large bump may halt radial dust drift entirely. The resulting increase in dust concentration may trigger planetesimal formation by the streaming instability (SI), as the SI itself requires some initial concentration of dust. 
Here we present the first 3D simulations of planetesimal formation in the presence of a pressure bump modeled specifically after those seen by ALMA. We place a pressure bump at the center of a large 3D shearing box, along with an initial solid-to-gas ratio of $Z = 0.01$, and we include both particle back-reaction and particle self-gravity. We consider mm-sized and cm-sized particles separately. For simulations with cm-sized particles, we find that even a small pressure bump leads to the formation of planetesimals via the streaming instability; a pressure bump does {\it not} need to fully halt radial particle drift for the SI to become efficient. Furthermore, pure gravitational collapse via concentration in pressure bumps (such as would occur at sufficiently high concentrations and without the streaming instability) is not responsible for planetesimal formation. For mm-sized particles, we find tentative evidence that planetesimal formation does not occur. If this result is confirmed at higher resolution, it could put strong constraints on where planetesimals can form. Ultimately, our results show that for cm particles planetesimal formation in pressure bumps is extremely robust.
\end{abstract}



\section{Introduction} 
\label{sec:intro}

One of the major open questions in planet formation theory is how planetesimals (1--100 km-size bodies) form out of mm--cm sized dust grains and pebbles. While collisions of $\mu$m--mm size silicate grains generally lead to sticking, once particles reach mm-cm sizes and collisions speeds reach $\Delta v_{\rm crit} \sim 1 \ {\rm m \ s^{-1}}$, collisions typically lead to bouncing or fragmentation \citep[e.g.][]{Guettler2010,Zsom10,Weidling2012,Kothe2013}. Given that typical turbulent velocities inside a protoplanetary disk are much larger than $\Delta v_{\rm crit}$ \citep{Ormel07}, planetesimals cannot grow by sticking of silicate particles. Some authors have suggested that icy aggregates are more sticky and can grow to larger sizes \citep{Wada2009,Gundlach2015}, but recent experiments suggest that this is only true within a very narrow temperature range of 175--200 K \citep{Musiolik2019}. For particles whose growth is not limited by bouncing or fragmentation, competition between particle growth and rapid radial drift also conspires to limit particles to the mm--cm size range, depending on the particle's location in the disk \citep{Weidenschilling77b,Birnstiel12}.

A promising mechanism to circumvent these barriers emerges when one accounts for both the aerodynamic drag and the momentum feedback onto the gas from the particles. The streaming instability (SI) is a radial convergence of particle drift that begins with a linear growth phrase \citep{Youdin05,Youdin07a,Squire_2020}, followed by a non-linear phase that causes a rapid concentration of particles into mostly axisymmetric filaments with a greatly enhanced density \cite[e.g.,][]{Johansen07a,Johansen07b,Bai10c,Li18,Abod19}. If the particle density exceeds the Roche density,

\begin{equation}
    \rho_{\rm roche} = \frac{9\Omega^2}{4\pi G},
\end{equation}
where $\Omega$ is the orbital frequency, then self-gravity between particles overpowers tidal forces. The resulting gravitationally bound clumps form with properties similar to Solar System asteroids and Kuiper Belt Objects \citep{Johansen07a,Johansen12,Johansen15,Simon16a,Simon17,Schafer17,Abod19,Nesvorny_2019,Li19}; planetesimals are born.

For the SI to operate, however, requires that the solid-to-gas ratio, $Z = \Sigma_{\rm solid} / \Sigma_{\rm gas}$, be sufficiently large.\footnote{The SI is most effective when all the solids are in cm-dm size pebbles. Since in all our simulations we used a single particle size, we will not distinguish between a solid-to-gas ratio and a ``pebble-to-gas'' ratio.} The critical $Z$ value needed to trigger the SI depends on the size of the particle stopping time, but in general $Z_{\rm crit} > 0.02$, and $Z_{\rm crit}$ rapidly increases for smaller particles. More quantitatively and under reasonable assumptions for the disk conditions at 1 AU, $Z_{\rm crit} \sim 0.02$ is needed for the SI to produce filaments for meter-size particles, and $\sim 0.03-0.04$ for small mm-size grains \citep{Carrera15,Yang_2017}, though it is worth noting \citet{Gerbig_2020} have recently argued that $Z_{\rm crit}$ is also proportional to the Toomre $Q$ parameter \citep{Toomre64} so that $Z_{\rm crit}$ is lower for young massive disks. For reasonable protoplanetary disk properties, however, such conditions are not regularly satisfied, and thus, a mechanism is required to enhance $Z$, either globally \citep[e.g., through photoevaporation of disk gas;][]{Carrera_2017}, or locally via dust pile-ups at the snowline \citep[e.g.][]{Ida16,Schoonenberg17,Drazkowska17} or local pressure bumps (the focus of this paper).

In recent years, the Atacama Large Millimeter/sub-millimeter Array (ALMA) has revealed a vast diversity of structures in nearby protoplanetary disks. Perhaps the most salient feature is a series of axisymmetric rings observed in the continuum emission associated with dust grains \citep{Alma15,Andrews_2018}. These radial concentrations of dust are thought to be caused by axisymmetric enhancements in gas pressure, or ``pressure bumps'', which may reduce or perhaps even reverse the direction of radial drift of solid particles \citep{Whipple72}. These structures may be the key ingredient required to enhance $Z$ to sufficient values such that the SI is activated and planetesimals can form. This is especially true if planetesimal formation happens sufficiently early such that photoevaporation is quite unlikely to be the relevant mechanism for increasing $Z$ and kick-starting the SI (see arguments in \citealt{Carrera_2017}). 

As discussed earlier, the key threshold that determines whether planetesimals form is that the mutual gravitational attraction between particles overpowers other forces. In the absence of significant velocity dispersion of these particles, gravitational collapse occurs when the enhanced density associated with particle clumping reaches the Roche density. The SI is perhaps the most promising route towards this critical collapse phase, but it may not be the only route. If solid particles can be sufficiently concentrated via other means, then the Roche density might be reached without the SI. As part of our investigation, we will explore whether pressure bumps that enhance $Z$ enough to trigger the SI may in fact by-pass the SI altogether and produce planetesimals via gravitational instability (GI) within the pressure bump. 

In addition to the question of {\it how} planetesimals form in pressure bumps (if at all), there are a number of other related questions we need to address in order build a more complete understanding of planetesimal formation.  First, do particles need to be {\it trapped} in a pressure bump (e.g., their radial drift halted), or will a weaker, transient enhancement in particle density as particles move through the bump suffice in producing planetesimals?  Answering this question will be key to understanding how robust planetesimal formation is and which bump-inducing mechanisms (e.g., planets, magnetically induced zonal flows; \citealt{Johansen09a}) are likely to form planetesimals. 

Second, where in relation to the pressure bump do planetesimals form? Do they form at the point of minimum radial drift, or elsewhere? If there is a pressure trap, at the exact center of the trap there is no headwind and the SI cannot be active (though the GI could be). Addressing this question will be crucial towards further developing planet formation models that make assumptions as to when and where planetesimals may form with respect to pressure bumps (e.g., \citealt{Stammler19,Eriksson20}). 

Finally, momentum feedback from particles within the pressure bump will influence the shape and structure of the bump in ways that are not yet clear. In turn, the potential deformation of the bump will likely itself influence the planetesimals that do form as a result.  This last question was first addressed in the work of \cite{Taki16}; they performed a 2D (radial-vertical) simulation of the vicinity of a radial pressure bump, and found that the pressure bump is completely deformed by the particle back-reaction.  They further found that the direct collapse of particles into planetesimals (i.e, by pure GI) is inhibited by this bump deformation, whereas the SI was active. Some key limitations of this work include the omission of stellar vertical gravity, particle self-gravity, and an azimuthal component to the simulations. \citet{Onishi17} corrected the first problem with a new 2D simulation that included vertical gravity. Because they allow dust particles to sediment, they found that the back-reaction is only a significant force in the thin dust layer at the midplane, where the particle density is high, and the majority of the pressure bump is largely unaffected. 

A common feature of \citet{Taki16} and \citet{Onishi17} is that the mechanism responsible for creating the pressure bump is assumed to be no longer active.  In contrast, our investigation is more focused on the scenario where the external force that created the bump is still actively reinforcing the bump; though we do include some simulations where there is no external reinforcement, and for these cases, we will make a direct comparison with previous work.  

Our paper is organized as follows. In \S\ref{sec:background} we give a brief review on radial drift, particularly within the context of pressure bumps. Then, in \S\ref{sec:methods} we summarize the numerical algorithms implemented in the {\sc Athena} code, followed by a description of the experimental setup and initial conditions in \S\ref{sec:setup}. Our results are presented in \S\ref{sec:results}. In \S\ref{sec:uncertainties} we discuss model uncertainties. Finally, we summarize and conclude in \S\ref{sec:conclusions}.

\section{Review of Radial Drift and Pressure Bumps}
\label{sec:background}

Solid particles in the disk experience aerodynamic drag. The stopping time for a particle with mass $m$ and material density $\rhos$ is given by

\begin{equation}
    \label{eqn:tstop}
    \tstop = \frac{m \, v_{\rm rel}}{F_{\rm drag}}
           = \frac{\rhos\, a}{\rho\, \cs} \sqrt{\frac{\pi}{8}},
\end{equation}
where $\rho$ is the gas density, $a$ is the particle radius, $\rhos$ is density of the solid material, and $\cs$ is the isothermal sound speed. The stopping time is typically expressed as the dimensionless Stokes number $\tau \equiv \tstop \Omega$, where $\Omega$ is the Keplerian frequency. The dominant Stokes number in a disk is set by various growth barriers \citep{Birnstiel12}. It turns out that in our simulations the particle size should be in the fragmentation-limited regime,

\begin{equation}
    \tau_{\rm frag} \approx \frac{v_{\rm frag}^2}{\alpha \; \cs^2},
\end{equation}

\noindent
where $v_{\rm frag}$ is the velocity at which particle collisions lead to fragmentation. For a fragmentation speed of $v_{\rm frag} \sim 1-2{\rm m} {\rm s}^{-1}$, low turbulence ($\alpha = 10^{-4}$), and $\cs \approx 370 {\rm m} {\rm s}^{-1}$ from our disk model (described in section \ref{sec:setup:disk}), we get $\tau_{\rm frag} \sim 0.07-0.3$.  The cm (mm) sized particles that we choose for our simulations (see Section~\ref{sec:setup:init}) correspond to $\tau \approx 0.12$ (0.012); thus our chosen particle sizes are consistent with this fragmentation limit.

The mm-cm size rage is also consistent with recent dust coagulation models in the vicinity of a pressure bump. In \citet{Stammler19}, the particle sizes are drift-limited in the vicinity of the pressure bump and only grow to the fragmentation limit at the center of the bump. While their pressure bump is not an exact analogue of ours (e.g. different shape, different location) it is worth noting that near the bump the typical particle size goes from around 1mm near the peak to $\sim$2cm at the peak.

For a locally isothermal disk (i.e., one with $\cs$ only dependent on the radial direction), hydrostatic equilibrium dictates that the vertical gas density profile must follow a Gaussian distribution with scale height $H = \cs/\Omega$.

The gas also experiences pressure support in the radial direction, as a result of the global temperature gradient. As a result, the gas orbits at a slightly sub-Keplerian speed. The difference between the Keplerian speed $\vk$, and the azimuthal speed of the gas $u_\phi$, results in solid particles experiencing a small headwind,

\begin{equation}
    \label{eqn:Delta_v}
    \Delta v \equiv \vk - u_\phi = \eta \vk
\end{equation}
where $\eta$ is given by \citet{Nakagawa86} as,

\begin{equation}
    \label{eqn:eta}
    \eta = - \frac{1}{2} \left( \frac{\cs}{\vk} \right)^2 \frac{d \ln P}{d \ln r}.
\end{equation}
For scale-free numerical simulations it is helpful to scale the headwind $\Delta v$ by the sound speed

\begin{equation}
    \label{eqn:PI}
    \Pi \equiv \frac{\Delta v}{\cs}
        = - \frac{1}{2} \left( \frac{\cs}{\vk} \right) \frac{d \ln P}{d \ln r}
\end{equation}

As solid particles experience a headwind, aerodynamic drag leads to the gradual loss of angular momentum, causing the solids to gradually drift toward the star. The rate of radial drift is 

\begin{equation}
    \label{eqn:vdrift}
    v_{\rm drift} = - \frac{2\,\Delta v}{\tau + \tau^{-1}}
                  = - \frac{2\,\eta\,\vk}{\tau + \tau^{-1}}.
\end{equation}

Note that the rate of radial drift is proportional to the logarithmic pressure gradient,

\begin{equation}
    v_{\rm drift} \propto \frac{d \ln P}{d \ln r}.
\end{equation}

If the disk has a pressure bump, the ``downstream'' side of the bump (i.e. toward the star) will have reduced $d \ln P / d \ln r$, leading to slower radial drift. This would create an over-density of particles, somewhat analogous to a traffic jam. If $d \ln P / d \ln r = 0$, particle drift stops entirely at that point, so that it becomes a particle trap.  Such concentration of particles may be a critical step for triggering planetesimal formation by the SI \citep{Carrera15,Yang_2017}. Regardless of the process that concentrates particles, once the particle density reaches the Roche density, the particle self-gravity overwhelms the Keplerian shear and the particles form a gravitationally bound clump, which will (upon further collapse) form planetesimals.

\section{Numerical methods}
\label{sec:methods}

For those readers familiar with our previous works,  feel free to read the following short text, skip the rest of Section~\ref{sec:methods}, and continue on to Section~\ref{sec:setup}: We conduct a series of local, shearing box simulations with the {\sc Athena} gas+particle code (ignoring magnetic fields and imposing no externally driven disk turbulence.) The gas is treated as a compressible, isothermal fluid, and the particles are treated via the super-particle approach.  Particle self-gravity is implemented using a particle-mesh approach with shearing-periodic radial boundary conditions.

\subsection{Hydrodynamic Solver}
\label{sec:methods:hydro}

We use the {\sc Athena} code in pure hydrodynamic mode with particle-gas interactions included and neglecting magnetic fields. We employ the local, shearing box approximation, in which we simulate a disk patch of sufficiently small size compared to the radial distance, $R_0$, from the central object that curvature effects can be neglected (though, see Section~\ref{sec:setup:resolution} for a description of why this may not be strictly true in our case). As such, the shearing box is a local Cartesian frame $(x,y,z)$, which is defined in terms of disk's cylindrical coordinate system $(R,\phi,z^\prime)$ as $x=(R-R_0)$, $y=R_0 \phi$, and $z = z^\prime$.  This box is co-rotating around the central object with an angular velocity $\Omega$, defined at the center of the box, $R_0$.  More details of the shearing box algorithm and its implementation can be found in \cite{Hawley95a} and (with respect to {\sc Athena}) \cite{Stone10}. Within this approximation, the equations of gas dynamics are: 

\begin{equation}
  \label{eqn:continuity_eqn}
  \frac{\partial \rho}{\partial t} + \del \cdot (\rho {\bmath u}) = 0,
\end{equation}

\begin{eqnarray}
  \label{eqn:momentum_eqn}
  \frac{\partial \rho {\bmath u}}{\partial t} 
     + \del \cdot \left(\rho {\bmath u}{\bmath u} + P {\bmath I} \right) 
     & = &  2 q \rho \Omega^2 {\bmath x} - \rho \Omega^2 {\bmath z} \nonumber \\
     &   & -2{\bmath \Omega} \times \rho {\bmath u} 
          + \rho_p \frac{{\bmath v}-{\bmath u}}{\tstop}.
\end{eqnarray}

\noindent
Where ${\bmath u}$ is the gas velocity and ${\bmath I}$ is the identity matrix. The shear parameter $q$ is defined as $q = -d\ln\Omega/d\ln r$, so that $q = 3/2$ for a Keplerian disk. From left to right, the source terms in equation~(\ref{eqn:momentum_eqn}) correspond to radial tidal forces (gravity and centrifugal), vertical gravity, the Coriolis force, and the particle momentum feedback onto the gas. In this last term, $\rho_p$ is the mass density of solid particles. The particle velocity vector is ${\bmath v}$, and $\tstop$ is the particle stopping time --- the timescale over which a particle will lose a factor of $e$ of its momentum due to gas drag. This feedback term is initially calculated at the particle locations and then distributed to the gas grid cells; we describe this mapping in more detail below. We also supplement these equations with an isothermal equation of state $P = \rho\cs^2$, where $\cs$ is the isothermal sound speed.

A second-order accurate Godunov flux-conservative method, coupled with the dimensionally unsplit corner transport upwind method of \cite{Colella90} and the third-order in space piece-wise parabolic method of \cite{Colella84} is used to solve the left-hand side of these equations (i.e., ignoring source terms).  A more detailed description and tests of these algorithms can be found in \cite{Gardiner05a}, \cite{Gardiner08}, and \cite{Stone08}.  Additional algorithms are used to integrate these equations within the shearing box approximation, thus handling the non-inertial source terms. These include orbital advection (the background Keplerian velocity is subtracted and integrated analytically; \citealt{Masset00,Johnson08,Johnson08b}) and Crank-Nicholson differencing, which is used to preserve epicyclic energy to machine precision. A detailed description of these algorithms, their implementation, and test problems are found in \cite{Stone10}.

{\sc Athena} includes a super-particle approach in which each super-particle is a statistical representation of a number of smaller particles. Super-particle $i$ (hereafter, ``particle" for simplicity) is governed by an equation of motion:

\begin{eqnarray}
  \label{eqn:particle_motion}
  \frac{d {\bmath v^\prime_i}}{dt} = 2\left( v^\prime_{iy} - \eta \vk \right)& & \Omega \hat{\bmath x} - \left(2 - q\right) v^\prime_{ix} \Omega \hat{\bmath y} \nonumber \\ 
& & - \Omega^2 z \hat{\bmath z} - \frac{{\bmath v^\prime_i} - {\bmath u^\prime}}{\tstop} + {\bmath F_{\rm g}}.
\end{eqnarray}

\noindent
In the above equation, the prime denotes a frame in which the background shear velocity has been subtracted, i.e., the orbital advection scheme mentioned above.  From left to right, the source terms are: radial acceleration of the particles due to the Coriolis effect, the gravitational and centrifugal forces, and radial drift; azimuthal motion due to the Coriolis effect; vertical motion due to the central star's gravity; gas drag; and the force due to particle self-gravity.

The $\eta\vk$ term is responsible for inward radial drift due to aerodynamic drag (see \S\ref{sec:background} and equation~(\ref{eqn:vdrift})); we follow \cite{Bai10a} and impose an inward force on the particles in the form of the $\eta\vk$ term. In practice, this means that the azimuthal velocities of particles and gas are shifted to slightly higher values (by  $\eta v_{\rm K}$) than would be present in a real disk.  Hence, we are allowed to maintain a Keplerian gas velocity profile as described above, and the particles are actually boosted to super-Keplerian speed.  This approach allows us to capture the essential physics of differential gas-particle motion.

Following \cite{Bai10a}, equation~(\ref{eqn:particle_motion}) is solved with a semi-implicit integration and a triangular shaped cloud (TSC) scheme that maps the particle momentum feedback to the grid cell centers (as mentioned above) and inversely interpolates the gas velocity to the particle locations (this interpolated quantity is ${\bmath u^\prime}$).  More details of this algorithm, in addition to test problems, can be found in \cite{Bai10a}.

\subsection{Particle Self-Gravity}
\label{sec:methods:self-gravity}

All of our simulations include particle self-gravity, with a corresponding force in the equation of motion represented by ${\bmath F_{\rm g}}$.  This term is found by solving Poisson's equation for particle self-gravity, and we use the same methods employed in \cite{Simon16a}. Briefly, we use the TSC scheme (mentioned above) to map the mass density of particles to grid cell centers. We shift the radial boundaries to be purely periodic and we use a Fast Fourier Transform (FFT) to solve the Poisson equation for the gravitational potential,

\begin{equation}
    \label{eqn:poisson}
    \nabla^2 \Phi = 4\,\pi\,G\,\rhop.
\end{equation}
Finally, the self-gravity force is ${\bmath F_{\rm g}} = - \nabla \Phi$ (see \citealt{Simon16a} and \citealt{Hawley95a} for more details). We employ open vertical boundaries for the potential, and as such, we apply a Green's function method to the Poisson equation in the vertical dimension  \cite[see][]{Koyama09,Simon16a}.   We then calculate the force due to self-gravity by applying a central finite difference, after which we interpolate this force (which is located at the grid cell centers) to the locations of the particles via TSC.  We have tested this algorithm in our previous work, \cite{Simon16a}, and more details can be found there.

\subsection{Boundary Conditions}
\label{sec:methods:bc}

The boundary conditions are the same for both the gas and particle components: shearing-periodic in the radial dimension \citep{Hawley95a}, purely periodic in azimuth, and a modified outflow boundary in the vertical dimension in which gas density is extrapolated into the ghost zones using an exponential function \citep{Simon11a,Li18}. This extrapolation has been shown to reduce gas mass loss and spurious effects near the vertical boundaries for vertically stratified shearing box simulations \citep{Simon11a}.

The modified outflow boundary condition will not entirely prevent gas mass loss along the vertical boundary, however, and to ensure that mass is globally conserved throughout our domain, we renormalize the gas density in every cell at every time step to keep the total gas mass constant. As for the particles component, we verify that no particles escape the simulation box through the vertical boundaries.

The gravitational potential has the same boundary conditions in the radial and azimuthal directions as the gas and particles. However, the vertical boundary conditions are open with the potential in the ghost zones calculated via a third order extrapolation.

\section{Simulation Setup}
\label{sec:setup}

Up until now, simulations of the SI have relied on small simulation domains and highly idealized initial conditions, such as a significantly enhanced solid-to-gas ratio. This was a result of computational cost and the need to explore a large and unfamiliar parameter space. 

Here we present large scale simulations with conditions grounded in recent observations of nearby circumstellar disks. Specifically, we model a large slice of a protoplanetary disk with an embedded axisymmetric pressure bump comparable to those responsible for observed dust rings.  Generally,
the gas structure that produces these dust rings is not well known.  However, \citet{Rosotti_2020} recently estimated the width of gas rings around a K-type (AS 209) and A-type (HD 163296) star, and they found $w/r \sim 7\%$ and $w/r \sim 22\%$, respectively, where $w$ is the standard deviation of a Gaussian density profile. 

To set up our bump properties, we assume a specific disk model that is loosely based on HL Tau (see below).  In particular, we consider a pressure bump located at $r_{\rm p} = 50$ AU, making it comparable to HL Tau's ring at 49 AU (B49) \citep{Alma15}. We choose a ring with FWHM = 12 AU, which is equivalent to $w/r \approx 10\%$, in approximate agreement with the gas ring widths mentioned above. 

\subsection{Disk Model}
\label{sec:setup:disk}

For our disk model, we assume a stellar mass of $M_\star = 1 M_\sun$ and a disk mass of $M_{\rm disk} = 0.09 M_\sun$. We model the disk surface density via a simple power law, 

\begin{equation}
  \label{eqn:sigma}
  \Sigma(r) = \frac{M_{\rm disk}}{2\pi r_c} r^{-1},
\end{equation}
where $r$ is the distance to the star, and $r_c = 200$ AU. We also assume a simple power law dependence for the gas temperature, consistent with an optically thin disk

\begin{equation}
  \label{eqn:temperature}
  T(r) = 280 {\rm K} \left(\frac{r}{{\rm AU}}\right)^{-1/2}.
\end{equation}
With this temperature profile, the disk aspect ratio is

\begin{equation}
    \label{eqn:aspect_ratio}
    \frac{H}{r} = \frac{\cs}{\vk} = 0.033 \left(\frac{r}{{\rm AU}}\right)^{1/4}.
\end{equation}
Next, using the isothermal approximation $P = \cs^2 \rho$, with $\cs \propto \sqrt{T}$, we obtain the background pressure profile,

\begin{equation}
    \label{eqn:background_pressure}
    P(r) \;=\;        \cs^2 \rho
      \;\propto\;  T \frac{\Sigma}{H}
      \;\propto\;  r^{-11/4}.
\end{equation}
Combining $d \ln P/d \ln r = - 11/4$ with equation~(\ref{eqn:aspect_ratio}) we obtain the headwind parameter,

\begin{equation}\label{eqn:PI_v2}
    \Pi(r) = 0.046 \left(\frac{r}{{\rm AU}}\right)^{1/4}
\end{equation}
and

\begin{equation}\label{eqn:eta_v2}
    \eta = 1.6 \times 10^{-3} \left(\frac{r}{{\rm AU}}\right)^{1/2}
\end{equation}

\subsection{Pressure Bump and Rossby Wave Instability}
\label{sec:setup:bump}

We model the pressure bump as a Gaussian in the gas density and pressure as follows

\begin{equation}
  \label{eqn:rho_bump}
  \rho(x,y,z) = \rho_0 \left[ 1 + A e^{\left(-x^2/2w^2\right)}\right] e^{\left(-z^2/2H^2\right)}
\end{equation}
where $A$ is the dimensionless amplitude of the bump, $w$ is the bump width, $H$ is the vertical gas scale height, and $x$ is the radial coordinate, centered at the peak of the pressure bump. Note that we include an additional Gaussian term in $z$ as well, which accounts for vertical hydrostatic equilibrium in an isothermal gas.

\begin{figure}[!ht]
    \centering
    \includegraphics[width=0.47\textwidth]{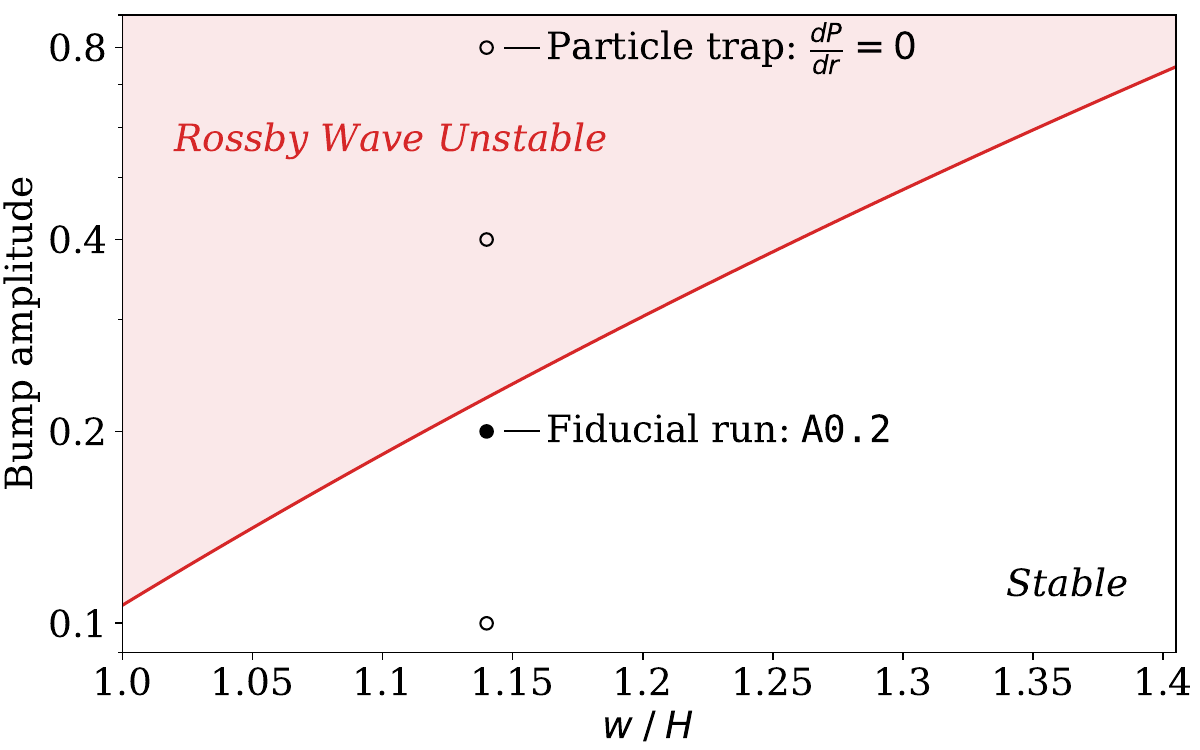}
    \caption{The red curve marks the maximum amplitude for a Gaussian pressure bump to be stable against the Rossby Wave Instability as derived via \cite{Ono16}. The plot assumes a disk aspect ratio of $H/r = 0.09$, which corresponds to our simulation setup. All our simulations (marked as circles) have a pressure bump at 50 AU with $w = 1.14 H$.}
    \label{fig:rossby}
\end{figure}

Since the amplitude of pressure bumps is not well-constrained by observations, we leave $A$ as a free parameter in our study.  However, a sufficiently large amplitude (for a given width) will render the system unstable to the Rossby Wave Instability (RWI; \citealt{Lovelace99}).\footnote{In principle, the Rayleigh instability could set in, but as discussed in \cite{Li00} and \cite{Ono16}, the stability condition for the RWI suffices to guarantee stability for the Rayleigh instability.} Assuming a constant surface density (i.e., no $r$ dependence on $\Sigma$), \citet{Ono16} derive a criterion for the maximum amplitude of a pressure bump that is stable against the RWI, as a function of the bump width. In the case of a Gaussian bump for an isothermal disk with $H/r = 0.1$ (case iv in their paper; most similar to our model) they find

\begin{equation}
    A_{\rm MS} = 1.06\times 10^5 \left( \frac{w}{r} \right)^{5.72}
\end{equation}

\noindent
where $A_{\rm MS}$ is the amplitude of marginal stability, $w$ is the bump's width (i.e., the standard deviation of the Gaussian), and $r$ is the semimajor axis of the bump. These coefficients are valid for $0.05 \le w/r \le 0.2$ (our model has $w/r \approx 0.1$). This stability criterion is shown in Figure \ref{fig:rossby}.

As noted earlier, our pressure bump has a FWHM of 12 AU and is placed at $r_{\rm p} = 50$ AU. At that semimajor axis, the disk scale height is $H = 4.5$ AU. Expressed in terms of the scale height, a Gaussian with a FWHM of 12 AU must have a standard deviation of $w = 1.14 H$. Figure \ref{fig:rossby} shows the stability criterion of \citet{Ono16} in terms of $w / H$. For $w = 1.14 H$, the amplitude of a marginally stable bump is $A = 0.226$. However, a bump that size will never be able to completely halt particle migration since the pressure gradient $dP/dr$ is always negative. A true particle trap ($dP/dr = 0$) requires a bump amplitude close to $A = 0.8$.  Despite the potential role of the RWI in disrupting a pressure bump with such an amplitude, exploring the role of such an amplitude on planetesimal formation will further our understanding of the relevant physics.  Furthermore, given the simplifications associated with the \cite{Ono16} work (e.g., a constant radial surface density profile), a bump that traps drifting particles may still be quite stable to RWI under different conditions.  

These considerations motivate the following exploration of the amplitude parameter:

\begin{itemize}
\item $A = 0.2$, since that is close to the largest bump amplitude consistent with Rossby wave stability.

\item $A = 0.8$ to include one run with a particle trap.

\item $A = 0.1$ and $0.4$ to explore parameter space. $A \in \{0.1, 0.2, 0.4, 0.8\}$ is uniform in log space.
\end{itemize}
The parameters for these bumps are shown in Figure \ref{fig:rossby}. 

To our knowledge, a study equivalent to \citet{Ono16} but with more realistic surface density structures has not been performed.  Thus, precisely which amplitudes in our parameter set are unstable to the RWI may change under more realistic conditions.

Finally, any long-lived pressure bump must be in geostrophic balance with the azimuthal flow. Integrating the momentum equation assuming such equilibrium and using equation~\ref{eqn:rho_bump}, we arrive at

\begin{equation}
  \label{eqn:uy_bump}
  u_y(x,y,z) = \frac{-A x\cs^2e^{\left(-x^2/2w^2\right)}}{ 2w^2\Omega\left[ 1 + A e^{\left(-x^2/2w^2\right)}\right]}
\end{equation}

\subsection{Newtonian Relaxation}
\label{sec:setup:T_reinf}

\begin{figure}
    \centering
    \includegraphics[width=0.47\textwidth]{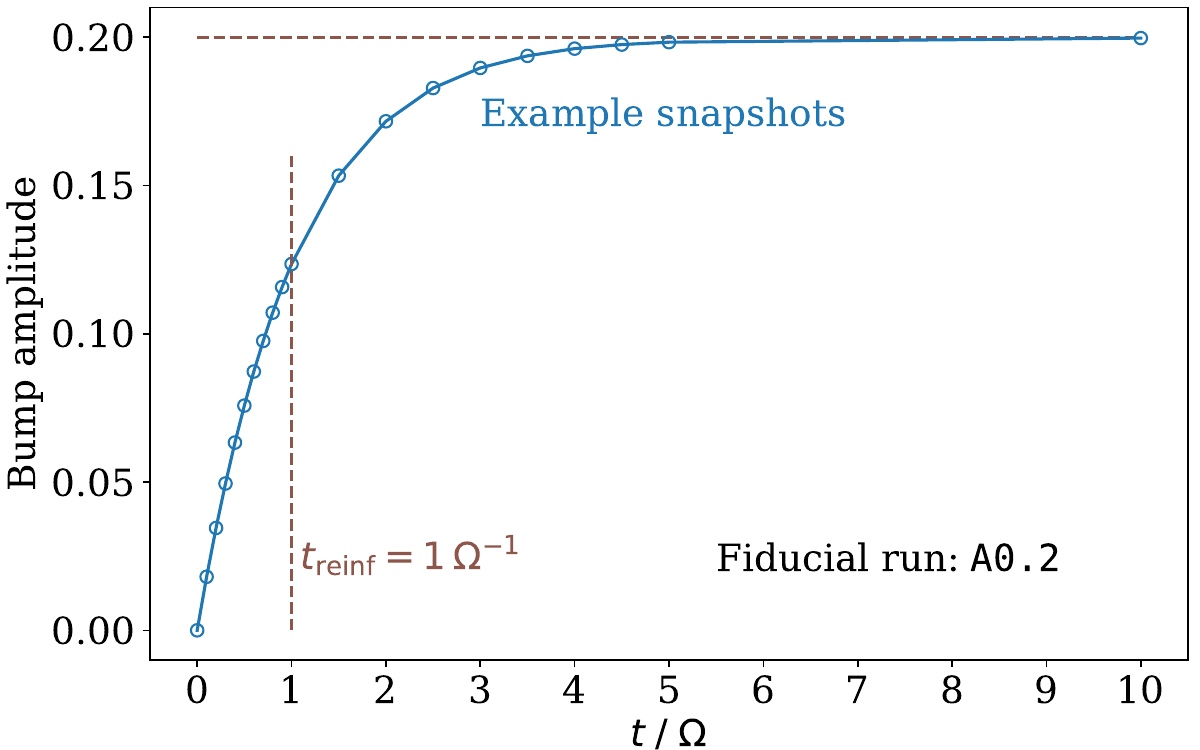}
    \caption{Amplitude of the pressure bump for our fiducial run, \texttt{A0.2}. Open circles mark simulation snapshots where the bump amplitude, $A = \max(\Sigma) / \min(\Sigma) - 1$, is calculated ($\Sigma$ is the gas column density). The amplitude converges toward $A = 0.2$ after a few reinforcement timescales.}
    \label{fig:amplitude}
\end{figure}

In the absence of particles, initializing the azimuthal gas speed according to equation~(\ref{eqn:uy_bump}) would be enough to ensure that the bump remains stable. However, particle feedback will gradually disrupt the bump unless there is an external force to reinforce it. For this work we assume that there is indeed some unspecified force (e.g., a planet or a magnetically induced zonal flow; \citealt{Johansen09a}) that reinforces the pressure bump on some reinforcement timescale $\treinf$. To simulate this process, we use Newtonian relaxation to adjust the radial profile of $\rho$ and $u_y$ 

\begin{eqnarray}
  \label{eqn:T_reinf}
  \Delta \rho &=& (\hat{\rho} - \rho)  \frac{\Delta t}{\treinf} \\
  \Delta u_y  &=& (\hat{u}_y  - u_y )  \frac{\Delta t}{\treinf}
\end{eqnarray}
where $\hat{\rho}$ and $\hat{u}_y$ denote the values in equations~(\ref{eqn:rho_bump}) and (\ref{eqn:uy_bump}) respectively. To illustrate how a pressure bump develops under our Newtonian relaxation scheme, we show (Figure \ref{fig:amplitude}) the evolution of the pressure bump amplitude,

\begin{equation}
    A(t) \equiv \frac{\max\left[\;\Sigma(t)\;\right]}{
                      \min\left[\;\Sigma(t)\;\right]} - 1
\end{equation}
in one of our simulations. The simulation begins with a uniform gas density. The target density profile $\hat{\rho}$ is a Gaussian bump with amplitude $A = 0.2$ and standard deviation $w = 1.14H$, and the reinforcement timescale is $\treinf = 1\;\Omegainv$. The figure shows the exponential convergence of the bump amplitude $A \rightarrow 0.2$.

\subsection{Domain Size and Resolution}
\label{sec:setup:resolution}

In order to encompass the entire width of the radial Gaussian and prevent spurious effects at the boundaries, we must choose a reasonably large radial domain size. We have found that $L_x = 9H$ is sufficiently large to avoid edge effects. In line with our previous works, we set the azimuthal extent of the domain to be $L_y = 0.2H$. We set the height to the domain to $L_z = 0.4H$ for runs with cm-sized particles and $L_z = 0.8H$ for mm-sized particles.\footnote{This larger vertical domain is required because the particle scale height will be higher for the mm-sized particles, thus leading to unacceptable mass loss from the domain.} This height ensures that no particles cross the vertical boundary and escape the domain.

Admittedly, $L_x = 9H$ is a long box that stretches the limits of the shearing box approximation. In particular, the length of this box normalized to its distance from the star is $L_x/2r_0 \approx 0.4$, clearly not satisfying the condition for the shearing box approximation: The error terms in the shearing box come from Hill's approximation. In the non-shearing frame we can write the equations of motion of particles as

\begin{eqnarray}
    \label{eqn:ddx_Hill}
    \ddot{x} - 2\Omega \dot{y}
         &=& \Omega^2 \left[ 1 + x - \frac{1+x}{r^3} \right]
                    + F_{x}\\
    \label{eqn:ddx_Hill2}
         &\approx& \Omega^2 (3x - 3x^2 + 4x^3 - \cdots) 
                    + F_{x}\\
    \label{eqn:ddy_Hill}
    \ddot{y} + 2\Omega \dot{x}
         &=& \Omega^2 \left[ y - \frac{y}{r^3} \right]
                    + F_{y}\\
         &\approx& \Omega^2 (3xy - 6x^2y + 10x^3y - \cdots)
                    + F_{y}
\end{eqnarray}
\noindent
where $(x,y)$ is the particle position normalized by $r_0$, $\vec{r} = (1+x,y)$, and $\vec{F}$ contains all other forces such as aerodynamic drag. For a shearing box like ours, where $L_x \gg L_y$, the largest error term is $E_x \approx - 3x^2\Omega^2$. Since the leading term in Equation (\ref{eqn:ddx_Hill2}) is $3x\Omega^2$, the relative error is in the order of $|E_x / 3x\Omega^2| \sim |x| \le L_x/2r_0$. Thus, for the shearing box to be a reasonable approximation,  $L_x/2r_0 \ll 1$. This is evidently not the case for our runs. However, we consider this an acceptable limitation because all of the relevant physics (aerodynamic drag, gas pressure, particle back-reaction, self-gravity) act on a very local scale ($\ll L_x$). In addition, the main alternative to a long shearing box (i.e., a global 3D simulation) is currently not feasible (both in terms of numerical resolution and code development).  

Given the large domain size, computational expense dictates a moderate resolution be employed. We use a standard resolution of 640 zones per $H$ for nearly all of our simulations. The resolution of 640 zones per $H$ is equivalent to the ``SI128" simulations of \cite{Simon16a}, which produced a reasonable number of planetesimals. Thus, with a standard domain size of $L_x\times L_y\times L_z = 9H\times0.2H\times0.4H$, our total resolution is $5760\times128\times256$. However, for the run with mm-size particles we decrease the resolution to 320 zones per $H$ in order to offset the cost of smaller timesteps needed for small-particle runs, and the greater vertical extent of the simulation domain. The total resolution for that run is $2880\times64\times256$. Finally, the total number of particles is the same as the number of grid cells --- $1.89 \times 10^8$ particles for our standard resolution runs.

\subsection{Initial Conditions and Parameters}
\label{sec:setup:init}

To recap, all runs have a pressure bump centered at $r_p = 50$ AU, with standard deviation $w = 1.14H$, and $H = 4.5$ AU (section \S\ref{sec:setup:bump}). At this location, the headwind parameter (equation (\ref{eqn:PI_v2})) is $\Pi = 0.12$. All simulations have a global dust/gas ratio of $Z = 0.01$ (a value comparable to that of the solar nebula) with no ab initio enhancement in the solid component. Simulations with an externally reinforced bump have a reinforcement timescale of $\treinf = 1\;\Omegainv$.

Our simulations use ``code units'' where $\cs = \Omega = H = 1$ define the units of length and time. For our disk model at 50 AU, the relative strength of tidal forces to self-gravity (i.e., the standard $\tilde{G}$ parameter in previous works, e.g., \citealt{Simon17,Abod19}) is,

\begin{equation}
    \tilde{G} \equiv \frac{4\pi G \rho_{\rm mid}}{\Omega^2} \approx 0.2,
\end{equation}
where $\rho_{\rm mid}$ is the midplane gas density of the unperturbed disk model. This corresponds to a gravitationally stable (in terms of the gas) disk with Toomre \citep{Toomre64} $Q \approx 8$. For our simulations, the midplane gas density is $\rho_{\rm mid} \approx \rho_0 \equiv 1$. In these units the Roche density is exactly $\rho_{\rm roche} = 9\Omega^2/4\pi G = 9\rho_{\rm mid}/\tilde{G} = 45$.

All simulations begin with a flat density profile in both the gas and solid components, with vertical stratification but no pressure bump,

\begin{equation}
  \label{eqn:rho_init}
  \left. \rho(x,y,z) \right|_{t=0} = 
        \left[ 1 + \frac{\rho_0 A w \sqrt{2\pi}}{L_x} \right]
        e^{\left(-z^2/2H^2\right)}
\end{equation}
Then the simulation is allowed to develop a pressure bump on its own. The scaling constant ensures that the midplane density $\rho(z=0)$ converges to a value close to $\rho_0$ at the edges of the box, and toward $\rho_0 (1 + A)$ in the middle of the box. 

One of the key parameters that controls the outcome of the SI is the dimensionless stopping time, or the Stokes number $\tau \equiv t_{\rm stop} \Omega$ (where $t_{\rm stop}$ is defined via Equation \ref{eqn:tstop}) \cite[e.g.][]{Carrera15}. Typical simulations of the SI assume that $\tau$ is constant, since particle and disk properties are assumed to be constant \cite[e.g.,][]{Simon16a,Abod19}. However, in the presence of a pressure bump, $\rho$ varies inside the simulation domain. So instead we assign the particle a fixed physical radius of either 1cm or 1mm and compute the Stokes number dynamically at every time step. For reference, a 1cm particle has a Stokes number of $\tau \sim 0.12$ away from the pressure bump in our setup. In all simulations, we choose only a single particle size for simplicity. Multiple particle sizes may very well have an effect on our results \cite[see][]{Krapp19}, and we leave the study of including more particle species for future work.

\begin{table*}[ht!]
\caption{We ran nine simulations. Our fiducial model, \texttt{A0.2}, has cm-size particles and a pressure bump with amplitude $A = 0.2$. Pressure bumps with amplitude $A \le 0.2$ are likely Rossby wave stable but we simulate bumps up to $A = 0.8$ because that is the amplitude needed to form a particle trap with $dP/dr = 0$. We also carry out a run in which we remove the particle-gas feedback, two runs without pressure bump reinforcement (section \S\ref{sec:setup:T_reinf}) so that the bump can dissipate by particle feedback, and two runs with mm-sized particles. Simulation \texttt{R0.8} was unintentionally run with a slightly larger particle size.}
\centering
\begin{tabular}{cccccccc}
Run  & Particle size 
     & Amplitude
     & RW stable?
     & Resolution 
     & Trap
     & Feedback
     & Reinforcement\\
\hline
\texttt{A0.1}        & 1 cm & 0.10 & \y & 640 / H & \n  & \y & \y \\
\texttt{A0.2}        & 1 cm & 0.20 & \y & 640 / H & \n  & \y & \y \\
\texttt{A0.4}        & 1 cm & 0.40 & \n & 640 / H & \n  & \y & \y \\
\texttt{A0.8}        & 1 cm & 0.80 & \n  & 640 / H & \y & \y & \y \\
\hline
\texttt{lores.1mm}   & 1 mm & 0.20 & \y & 320 / H & \n  & \y & \y \\
\texttt{hires.1mm}   & 1 mm & 0.20 & \y & 640 / H & \n  & \y & \y \\
\hline
\texttt{NoFeedback} & 1 cm & 0.80 & \n & 640 / H & \y & \n & \y \\
\texttt{R0.2}       & 1 cm & 0.20 & \y & 640 / H & \n & \y & \n \\
\texttt{R0.8}       & 1.6 cm & 0.80 & \n & 640 / H & \y & \y & \n \\
\end{tabular}
\label{tab:runs}
\end{table*}

\subsection{List of Simulations}
\label{sec:setup:summary}

Altogether, we conducted nine simulations (which are also summarized in Table~\ref{tab:runs}):

\begin{description}
  \item[\texttt{A0.2}] \hfill \\
        Our fiducial run, with a fixed particle size of 1 cm, a pressure bump with an amplitude of $A = 0.2$, and at full resolution (640 zones/$H$; see \S\ref{sec:setup:resolution}).
  \item[\texttt{A0.8}] \hfill \\
        A full resolution run with cm-sized particles and a pressure bump amplitude of $A = 0.8$. While this bump may be Rossy wave unstable (see \S\ref{sec:setup:bump}), it is important to quantify the effect of a particle trap ($d \ln P / d \ln r = 0$).
  \item[\texttt{A0.1}, \texttt{A0.4}] \hfill \\
        Two runs to complete the exploration of amplitude parameter space.
  \item[\texttt{hires.1mm}, \texttt{lores.1mm}] \hfill \\
        To investigate the effect of particle size, we run simulations with mm-sized particles. Due to the high computational cost, we cannot run long-term simulations of mm-sized particles at full 640/H resolution. Instead, we do a short simulation at full resolution and a long-term simulation at half-resolution (320/$H$).
  \item[\texttt{NoFeedback}] \hfill \\
        To test whether planetesimals form by the SI and not gravitational instability (GI) induced directly from particle concentration in the pressure trap, we run one simulation with the particle feedback (critical for the SI) turned off.
  \item[\texttt{R0.2}, \texttt{R0.8}] \hfill \\
        Lastly, we run two simulation that \textit{begin} with a fully formed bump in geostrophic balance, but with no external reinforcement (see \S\ref{sec:setup:T_reinf}). Then we measure how quickly the bump is destroyed by particle feedback, and whether planetesimals can form before the bump dissipates.
\end{description}

\section{Results}
\label{sec:results}

\subsection{Particle Density and the Roche Density}
\label{sec:results:d_max}

\begin{figure}
    \centering
    \includegraphics[width=0.47\textwidth]{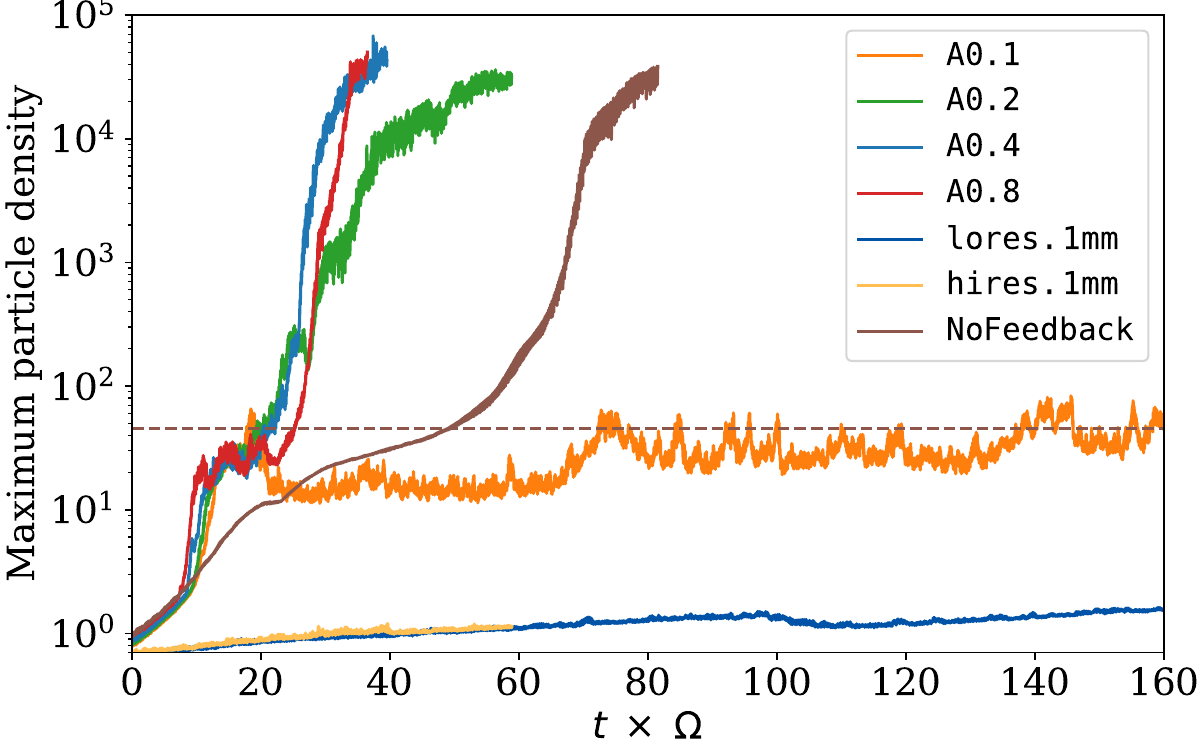}
    \caption{Maximum particle density $\rho_{\rm p,max}$ (in code units) versus time. The dashed line marks the Roche density $\rho_{\rm roche}$. Most models (\texttt{A0.2}--\texttt{A0.8}) go through three main stages: (1) initial sedimentation ($t \approx 0-10\;\Omegainv$), (2) the SI forms dense filaments that cross the Roche density, and (3) gravitational collapse of over-dense regions into particle clumps. All simulations with $A > 0.1$ form gravitationally bound clumps. Simulations with mm-size grains (\texttt{lores.1mm} and \texttt{hires.1mm}) have very slow density growth and may never reach the Roche density (see section \S \ref{sec:results:small}).}
    \label{fig:d_max}
\end{figure}

Figure \ref{fig:d_max} shows the maximum particle density as a function of time for most of our runs. All models with cm-size particles begin with an initial sedimentation phase that increases the maximum particle density $\rho_{p,\rm max}$ for the first $\Delta t \sim 10\;\Omegainv$ of evolution. At that moment particle feedback begins to alter the velocity of the gas, so the \texttt{NoFeedback} simulation separates from the other models. For models \texttt{A0.2}--\texttt{A0.8} it takes another $\Delta t \sim 10\;\Omegainv$ for the SI to form filaments with sufficient density to trigger self-gravity. The key result here is that all simulations with amplitude $A \ge 0.2$ and cm-size particles achieved particle densities several orders of magnitude greater than the Roche density. Since model \texttt{A0.8} is the only model with a particle trap (meaning that $\Delta v = 0$), we find that a particle trap is not needed to form planetesimals by the SI. Furthermore, a pressure bump that is Rossby-wave stable (again, according to the criterion of \citealt{Ono16}; model \texttt{A0.2}) is seen to trigger planetesimal formation.

\subsection{Comparison to a Simple Model}
\label{sec:results:naive}

\begin{figure}
    \centering
    \includegraphics[width=0.47\textwidth]{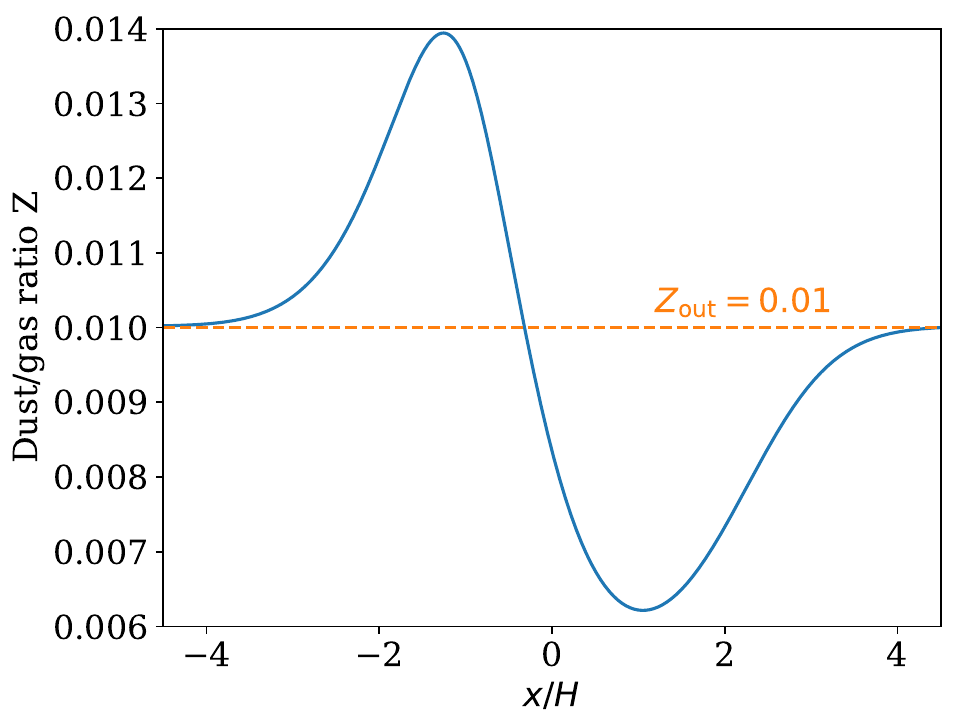}
    \caption{solid-to-gas ratio $Z$ implied by the simple steady state model of Equation \ref{eqn:particle_mass_flux} for a pressure bump with amplitude $A = 0.2$. The peak solid-to-gas ratio is $Z_{\rm max} = 0.0139$, independent of $\tau$. This is a borderline value for cm-size particles ($\tau \approx 0.12$) and too small to clump particles for mm-size particles \citep{Yang_2017}.}
    \label{fig:naive}
\end{figure}

To develop some intuition as to why planetesimal formation is still possible, even without trapping, we consider a pressure bump that reaches a steady state with a constant particle mass flux and has no particle trap,

\begin{equation}
    \label{eqn:particle_mass_flux}
    \dot{M}_{\rm p} = 2\pi r \Sigmap \vr = const,
\end{equation}
where $\Sigmap$ is the particle surface density and $\vr$ is the radial velocity. The pressure bump creates a localized reduction of $\vr$ on the downstream side of the bump, and a corresponding localized increase in $\Sigmap$. In effect, there is a traffic jam of particles in the region where the particle drift is slowed down, and this traffic jam leads to a local enhancement in the solid-to-gas ratio. Figure \ref{fig:naive} shows the solid-to-gas ratio $Z$ predicted by the steady state model for a pressure bump with $A = 0.2$. Note that, while $\vr$ and $\dot{M}_{\rm p}$ clearly depend on the particle Stokes number, the particle density ($Z \propto \Sigmap \propto \dot{M}_{\rm p} / \vr$) is independent of $\tau$. The steady state model gives a peak solid-to-gas ratio of $Z_{\rm max} = 0.0139$, which is nearly identical to the critical $Z_{\rm crit}$ predicted by \citet{Yang_2017} for cm-size particles ($\tau \approx 0.12$) and much smaller than the $Z_{\rm crit} = 0.025$ for mm-size particles.

In other words, the steady state model predicts that runs \texttt{A0.4} and \texttt{A0.8} should trigger strong clumping, that run \texttt{A0.2} is marginal, and that runs \texttt{A0.1} and \texttt{*.1mm} \textit{should not} produce strong clumping. Considering all the simplifications, the simple steady state model was surprisingly predictive.

\subsection{Time and Location of Planetesimal Formation}
\label{sec:results:time_and_place}

\begin{figure*}[t]
    \centering
    \includegraphics[width=0.85\textwidth]{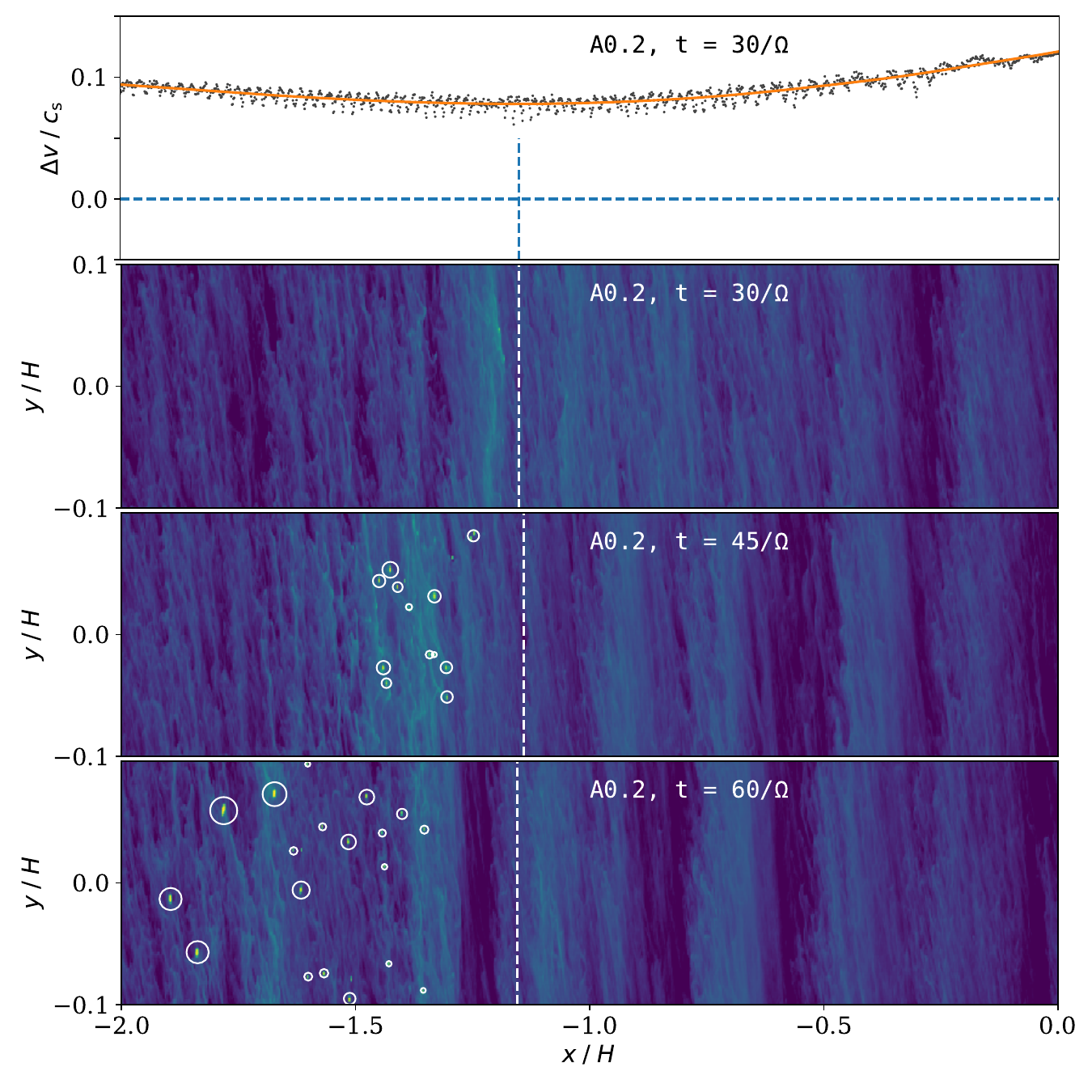}
    \raisebox{0.5\height}{\includegraphics[width=0.11\textwidth]{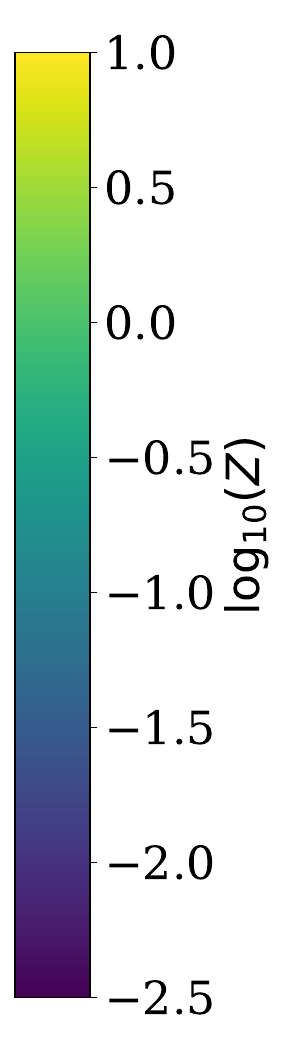}}
    \caption{Snapshots of model \texttt{A0.2} at the time when particle clumps form. \textit{Top:} The black dots show the azimuthally averaged value of the headwind at $t = 25/\Omega$. The orange line corresponds to the same data as the black dots, but averaged over radially nearby points. The vertical dashed line marks the location where the orange line reaches its minimum value. \textit{Bottom 3 plots:} Snapshots of the column dust/gas ratio $Z = \Sigmap / \Sigma$. White circles mark the location of bound clumps with $\rhop \gg \rho_{\rm roche}$.}
    \label{fig:formation_A02}
\end{figure*}

When planetesimal formation does occur, it is rapid, requiring only a few tens of $\Omegainv$. The run with no particle feedback takes twice as long to cross the Roche density as the equivalent run with feedback (i.e.~{\tt A0.8}). We shall return to the role of feedback in section \S\ref{sec:results:SI_vs_GI}.

Finally, the reader may notice that the simulations with no bump reinforcement (runs {\tt R0.2} and  {\tt R0.8}) are not shown in figure \ref{fig:d_max}. This is intentional. These runs begin with a fully formed pressure bump, which gives them a head start in planetesimal formation. We shall discuss these runs, and the effect of removing the bump reinforcement, in section \S\ref{sec:results:reinforcement}.

\begin{figure*}[t]
    \centering
    \includegraphics[width=0.85\textwidth]{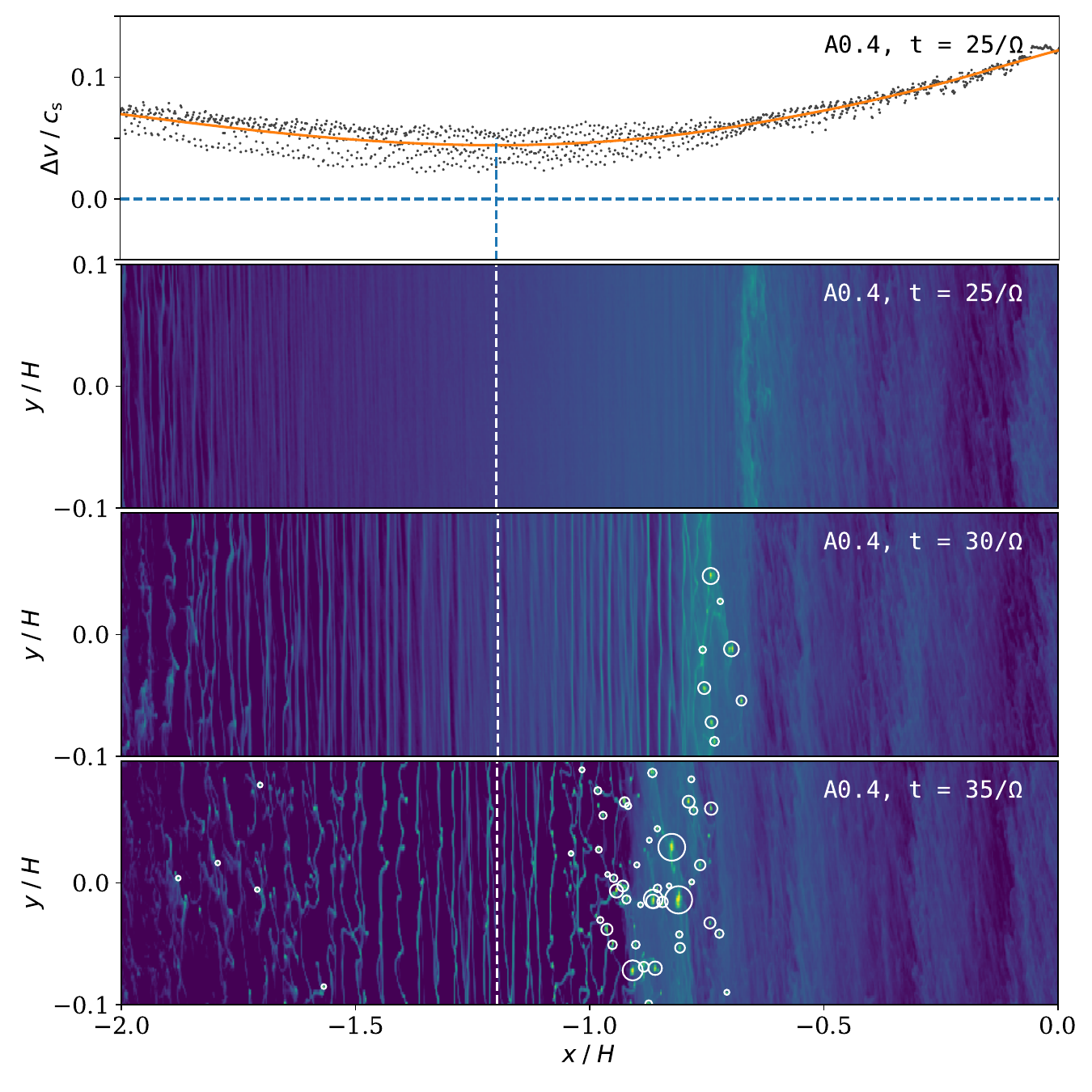}
    \raisebox{0.5\height}{\includegraphics[width=0.11\textwidth]{colorbar_vertical.pdf}}
    \caption{Snapshots of model \texttt{A0.4} at the time when particle clumps form. \textit{Top:} The black dots show the azimuthally averaged value of the headwind at $t = 25/\Omega$. The orange line corresponds to the same data as the black dots, but averaged over radially nearby points. The vertical dashed line marks the location where the orange line reaches its minimum value. \textit{Bottom 3 plots:} Snapshots of the column dust/gas ratio $Z = \Sigmap / \Sigma$. White circles mark the location of bound clumps with $\rhop \gg \rho_{\rm roche}$.}
    \label{fig:formation_A04}
\end{figure*}

\begin{figure*}[t]
    \centering
    \includegraphics[width=0.85\textwidth]{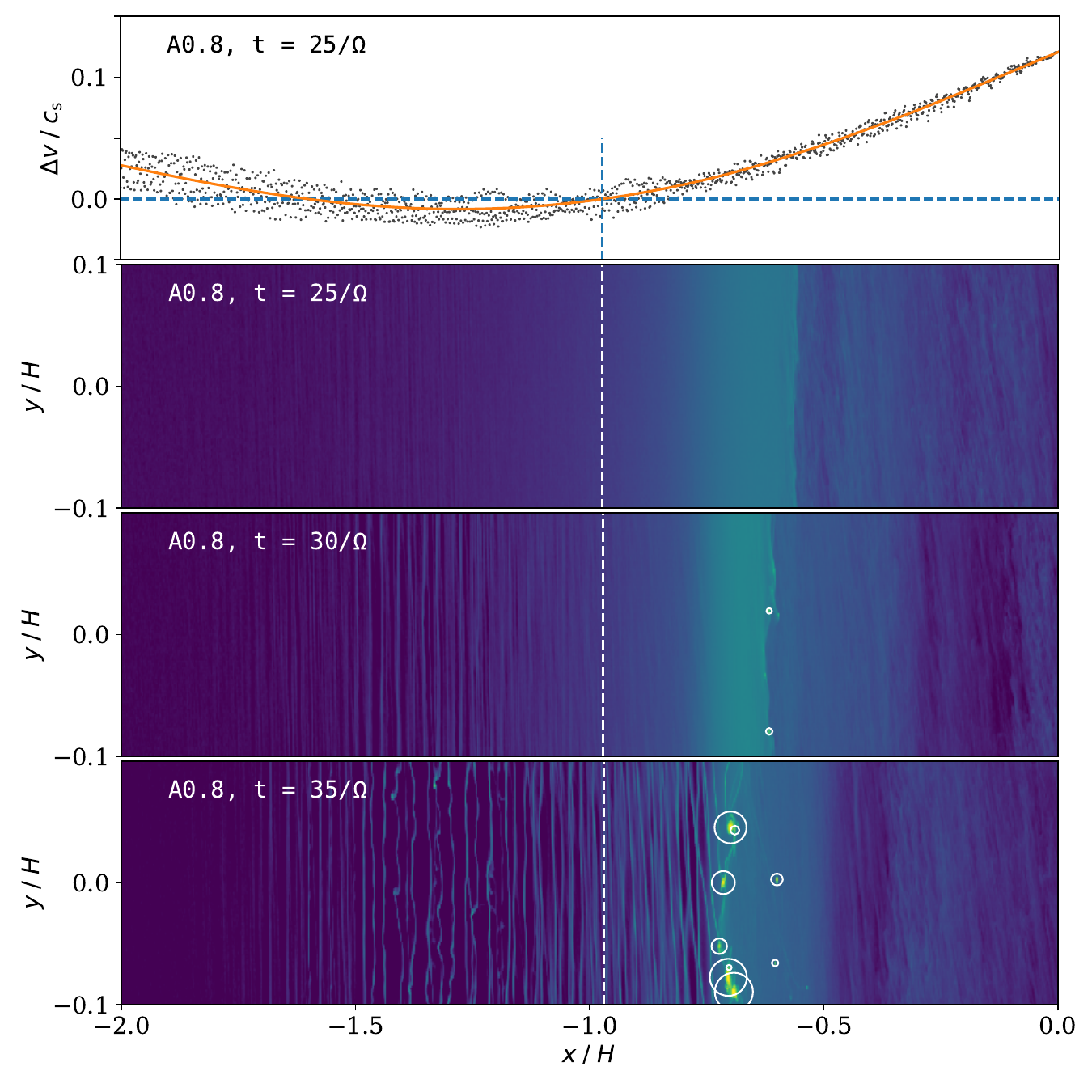}
    \raisebox{0.5\height}{\includegraphics[width=0.11\textwidth]{colorbar_vertical.pdf}}
    \caption{Snapshots of model \texttt{A0.8} at the time when particle clumps form. \textit{Top:} The black dots show the azimuthally averaged value of the headwind at $t = 25/\Omega$. The orange line corresponds to the same data as the black dots, but averaged over radially nearby points. The vertical dashed line marks the location where the orange line reaches its minimum value.\textit{Bottom 3 plots:} Snapshots of the column dust/gas ratio $Z = \Sigmap / \Sigma$. White circles mark the location of bound clumps with $\rhop \gg \rho_{\rm roche}$.}
    \label{fig:formation_A08}
\end{figure*}

Figures \ref{fig:formation_A02}--\ref{fig:formation_A08} show snapshots of simulations \texttt{A0.2}, \texttt{A0.4}, and \texttt{A0.8} at the time when bound particle clumps begin to form. The figures also show the headwind $\Delta v$, as well as the location of the particle trap ($\Delta v = 0$) for \texttt{A0.8}, or the location where $\Delta v$ reaches its minimum value for \texttt{A0.2} and \texttt{A0.4}. Some important takeaways include,

\begin{itemize}
\item Planetesimal formation is not restricted to a fixed point in space. The planetesimal forming filaments drift, and the particle clumps drift. In all cases, planetesimal formation is radially dispersed. Presumably, when clumps collapse into planetesimals those will stop drifting, but our simulations cannot resolve the final collapse. The collapse itself is likely to be much faster (tens of years) than the orbital timescale at 50 AU \citep{Jansson_2014,Nesvorny_2010}.

\item Even when a particle trap is present (run \texttt{A0.8}), planetesimal formation does not occur at the trap. In \texttt{A0.4} particle clumps form before the minimum $\Delta v$ and in run \texttt{A0.2} they form after. Compare this with \citet{Onishi17}, who found that planetesimals formed at the particle traps. However, their run had a lower $\Pi$ (0.05 vs 0.12 for our runs) so that a bump with $A = 0.2$ formed a particle trap. Evidently there is an interplay between $\Pi$ and $A$ and in principle planetesimal formation can occur on either side of the minimum $\Delta v$. The $\Delta v$ needed to trigger the SI is some value $> 0$; run \texttt{A0.4} must have reached this value before the minimum $\Delta v$.

\item In model \texttt{A0.2} planetesimal formation is a much slower, more gradual process. Notice that in figure \ref{fig:formation_A02} the snapshots are taken at intervals of $\Delta t = 15/\Omega$ while figures \ref{fig:formation_A04} and \ref{fig:formation_A08} have snapshots taken at intervals of $\Delta t = 5/\Omega$.

\item Figure \ref{fig:formation_A02} suggests that simulation \texttt{A0.2} might be a marginal case. The bound clumps form \textit{after} the minimum $\Delta v$. The densest filament forms at the location of ${\rm min}(\Delta v)$ (snapshot $t = 30/\Omega$) and then drifts before bound clumps form.

\item Model \texttt{A0.2} has ${\rm min}(\Delta v) \approx 0.08 \cs$ at $t = 30/\Omega$. If \texttt{A0.2} is indeed a marginal case, then $\Delta v \approx 0.08 \cs$ may be close to the critical value needed for a pressure bump to trigger planetesimal formation by the SI.

\end{itemize}

\subsection{Streaming or Gravitational Instability?}
\label{sec:results:SI_vs_GI}

\begin{figure*}
    \centering
    \includegraphics[width=0.85\textwidth]{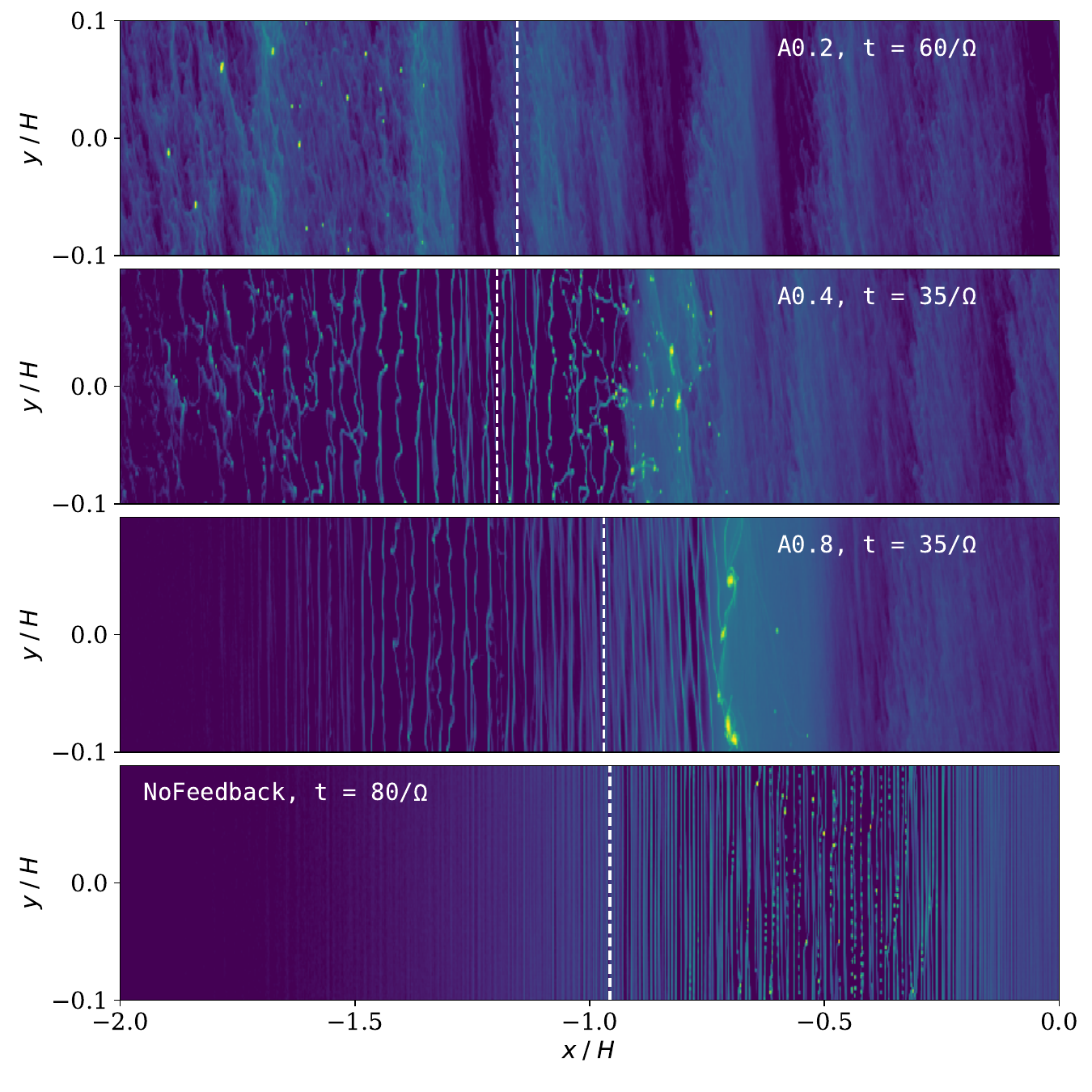}
    \raisebox{0.16\height}{\includegraphics[width=0.11\textwidth]{colorbar_vertical.pdf}}
    \caption{Snapshots of the column dust/gas ratio ($Z = \Sigmap / \Sigma$) at the time that bound particle clumps form in models \texttt{A0.2}--\texttt{A0.4} and \texttt{NoFeedback}. Notice the difference between the thick filaments formed in runs where the SI is active (\texttt{A0.2}--\texttt{A0.4}) versus a model where planetesimals can only form by GI. Notice also that, even with a particle trap, GI takes longer to form planetesimals than the SI. Therefore, we find that the SI (and not GI) is the mechanism responsible for forming planetesimals in runs \texttt{A0.2}--\texttt{A0.8}.}
    \label{fig:compare_A02_NFB}
\end{figure*}

Our next experiment is designed to confirm whether the self-gravitating particle clumps in our simulations were truly the result of the SI, or whether GI induced by pure concentration in the bump is responsible. The SI is a radial convergence of particle drift caused by the particle-on-gas feedback (see section \S\ref{sec:intro}). Therefore, we ran a simulation, {\tt NoFeedback}, where we completely removed the particle feedback. This model has a bump amplitude of $A = 0.8$, so that it has a true particle trap (i.e. headwind $\Delta v = 0$) to maximize the particle concentration.

Figures \ref{fig:d_max} and \ref{fig:compare_A02_NFB} show that removing the particle feedback significantly delays the particle accumulation, as well as planetesimal formation, even in a simulation designed to trap particles. The panels of figure \ref{fig:compare_A02_NFB} show 2D plots from runs {\tt A0.2}--{\tt A0.8} and {\tt NoFeedback} at a time near their peak particle density. The differences in the filamentary structure show that the {\tt NoFeedback} run is dominated by very different physical processes. Runs {\tt A0.2}--{\tt A0.8} show the wide filaments with complex sub-structure that are characteristic of the SI. The filaments in run \texttt{NoFeedback} are much narrower, far more closely spaced, and do not have nearly as much sub-structure. Even if we compare \texttt{NoFeedback} to {\tt A0.8} (the run with trapping {\it and} feedback), there is a noticeable difference in the structure and sizes of formed clumps. Ultimately, these considerations strongly imply that it is the SI and not pure GI due to concentration that gives rise to planetesimal formation in pressure bumps.

\subsection{Small Particles and Resolution Limit}
\label{sec:results:small}

\begin{figure}
    \centering
    \includegraphics[width=0.47\textwidth]{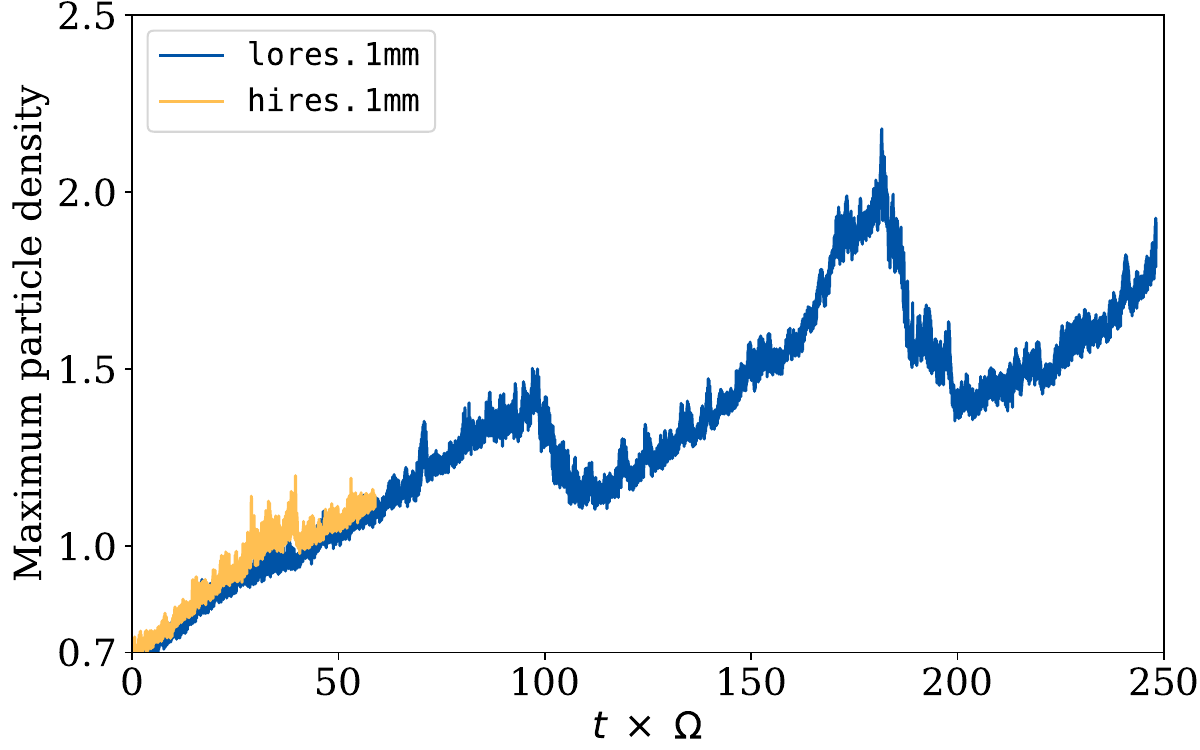}
    \caption{Maximum particle density (in code units) versus time for the two runs with mm-sized particles. The high-resolution run closely follows the low-resolution one, which displays slow growth. At the present growth rate, mm-size runs may not reach the Roche density until $t \sim 9,000\;\Omegainv$, which is much longer than the particle drift timescale.}
    \label{fig:d_max_zoom}
\end{figure}

In the case of mm-sized particles it appears that it may not be possible for a Rossby-wave stable bump to trigger planetesimal formation by the SI. Figure \ref{fig:d_max_zoom} zooms into the two runs with mm-size particles and extends the range of the time integration. Because the high-resolution run closely follows the low-resolution run, it appears that at least the bulk behaviour of the mm-size particles is mostly resolved. We extended \texttt{lores.1mm} to $t = 250 \Omegainv$ and only observed slow gradual growth of particle density. Simple linear extrapolation suggests that particles will reach the Roche density ($\rho_{\rm roche} = 45$) at around $t_{\rm roche} \sim 9,000\;\Omegainv$. For comparison, the radial drift timescale (using equations (\ref{eqn:vdrift}) and (\ref{eqn:eta_v2})) is

\begin{equation}
    t_{\rm drift} \sim \frac{v_{\rm drift}}{r}
                  \sim 2,270 \Omegainv.
\end{equation}

Since $t_{\rm drift} \ll t_{\rm roche}$, these runs suggest that mm-size particles may never reach the Roche density. The most important question now is whether this result reflects physical reality, or is merely a numerical limitation of our simulations. Specifically, it is possible that neither of these runs had a sufficiently high resolution to resolve the fastest growing modes of the SI. In the SI, the fastest growing modes are those that satisfy the epicyclic resonant condition \citep{Squire_2020}

\begin{equation}
    \label{eqn:si_fast}
    \mathbf{k} \cdot \mathbf{w}_s = \hat{k}_z \Omega,
\end{equation}
where $\mathbf{k} = (k_x, 0, k_z)$ is the wavenumber, $\mathbf{w}_s$ is the dust drift velocity with respect to the gas, and $\hat{k}_z = k_z/k$. Since $\mathbf{w_s} \approx -2 \tau \eta \; \vk \; \mathbf{\hat{x}}$ \citep{Nakagawa86}, the fastest growing mode has wavelength

\begin{equation}
    \lambda \approx -4\pi\tau \eta\;r\;\frac{k_x}{k_z}.
\end{equation}
For our disk model at 50 AU, and assuming the minimum value of $\eta$ in the bump, we get $\lambda/H \approx 0.72\tau|k_x/k_z|$ for $A = 0.2$. Therefore, the resolution required to resolve the fastest growing mode of the SI is inversely proportional to the particle size. Our \texttt{hires.1mm} model has a resolution of $H/\Delta x = 640$ and \texttt{lores.1mm} has $H/\Delta x = 320$. If we let $|k_x/k_z| = 1$, then $\lambda/\Delta x \approx 3 - 6$. In other words, we should be able to resolve the fastest growing mode if $|k_z| \le |k_x|$, though given the low value of $\lambda/\Delta x$, this could be marginal.

It is worth keeping in mind that equation~(\ref{eqn:si_fast}) only applies for local solid-to-gas ratios less than unity. We find that 98.9\% of the particle mass at the end of the \texttt{lores.1mm} is in regions where $\rhop / \rho < 1$. Thus, the majority of our simulation domain should be susceptible to the $\rhop / \rho < 1$ limit of the streaming instability \citep{Squire_2020}.  Furthermore,
as pointed out in both \cite{Squire_2020} and the original work by \cite{Youdin05}, the condition $\rhop / \rho > 1$ is generally seen as the criterion for the existence of faster growing SI modes (i.e., growth rates dramatically increase above the $\rhop / \rho = 1$ value).  Whether it is possible for this fast-growing SI to be fully manifested in only 1.1\% of our simulation (in terms of solid-to-gas ratio) is not entirely clear.   Regardless, the fact that most of the particle mass in \texttt{lores.1mm} falls within the $\rhop / \rho < 1$ regime and that the fastest unstable SI modes are resolved by $\sim 3$--6 grid cells suggests that we might be able to see at least some indication of exponential SI growth; yet, we do not.

\begin{figure}
    \centering
    \includegraphics[width=0.47\textwidth]{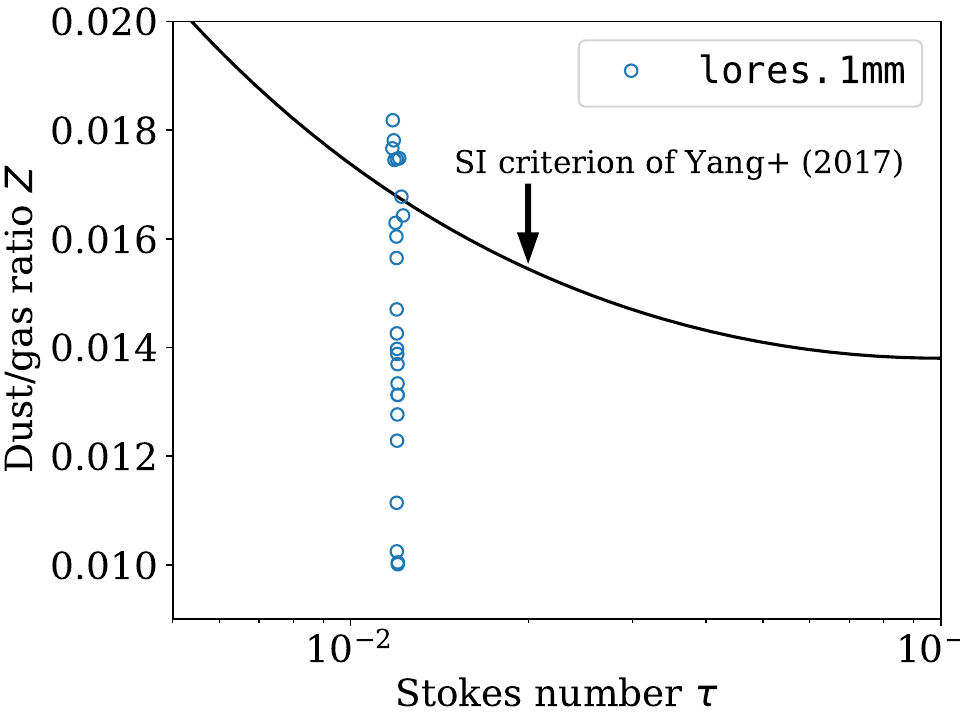}
    \caption{The black curve shows the SI criterion of \citet{Yang_2017}; disk conditions above the black curve are thought to be conducive to the SI. The blue circles mark the average dust to gas ratio $Z$ (calculated as described in the main text) and Stokes number $\tau$ for run \texttt{lores.1mm}, computed in intervals of $10\;\Omegainv$.}
    \label{fig:SI_criterion}
\end{figure}

 We have also calculated the local-particle-density-weighted average of $Z$ and $\tau$ for \texttt{lores.1mm}, as shown in Fig.~\ref{fig:SI_criterion}.  At first blush, it appears that the simulation has reached the region where it should be unstable to the SI (according to the criterion of \citealt{Yang_2017}). However, the SI criterion of \cite{Yang_2017} was computed with a headwind parameter of $\Pi = 0.05$, whereas our simulations have $\Pi = 0.12$. Lower headwind is known to facilitate particle concentration by the SI \citep{Bai10b}, and thus, it is possible
 that the \cite{Yang_2017} criterion would be pushed towards higher $Z$ values for our larger $\Pi$ value.  If indeed, the critical $Z$ for $\Pi = 0.12$ is larger than that reached in our simulation, that would explain the lack of strong clumping (and thus planetesimal formation).
 
 Ultimately, to resolve these issues of the SI and its ability to produce planetesimals, we will need to carry out higher resolution simulations that explore a larger area of the parameter space.  Given the rich physics associated with cm-sized particles and pressure bumps, which is the focus of this paper, we defer such work to a future publication.

\subsection{Resilience of Particle Filaments}
\label{sec:results:resilience}

\begin{figure*}[t]
    \centering
    \includegraphics[width=0.85\textwidth]{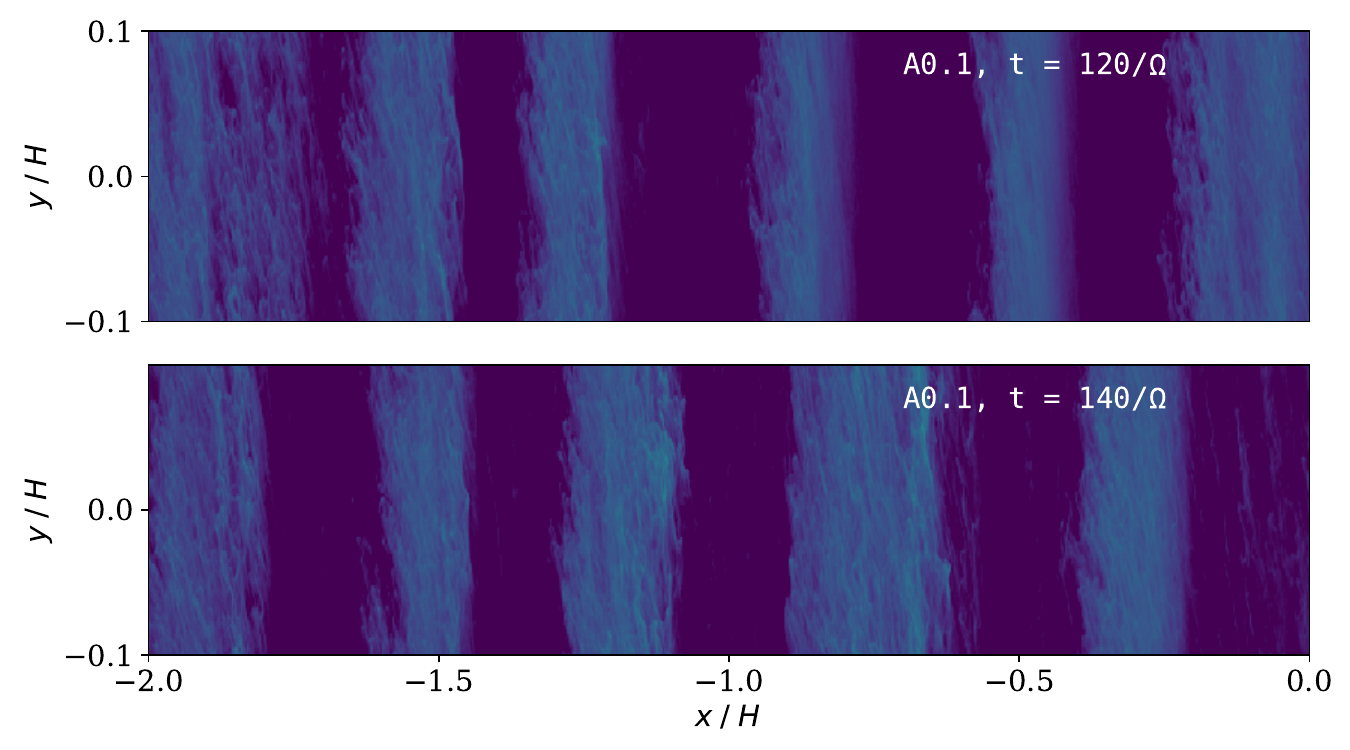}
    \raisebox{0.1\height}{\includegraphics[width=0.11\textwidth]{colorbar_vertical.pdf}}
    \caption{Snapshots of the column dust/gas ratio $Z = \Sigmap / \Sigma$ in model \texttt{A0.1}. The simulation does not form planetesimals but particle filaments are clearly visible. Filaments like these are seen across the entire simulation box even in models that do not form planetesimals (see Figure \ref{fig:filaments}).}
    \label{fig:compare_A01}
\end{figure*}

\begin{figure}
    \centering
    \includegraphics[width=0.47\textwidth]{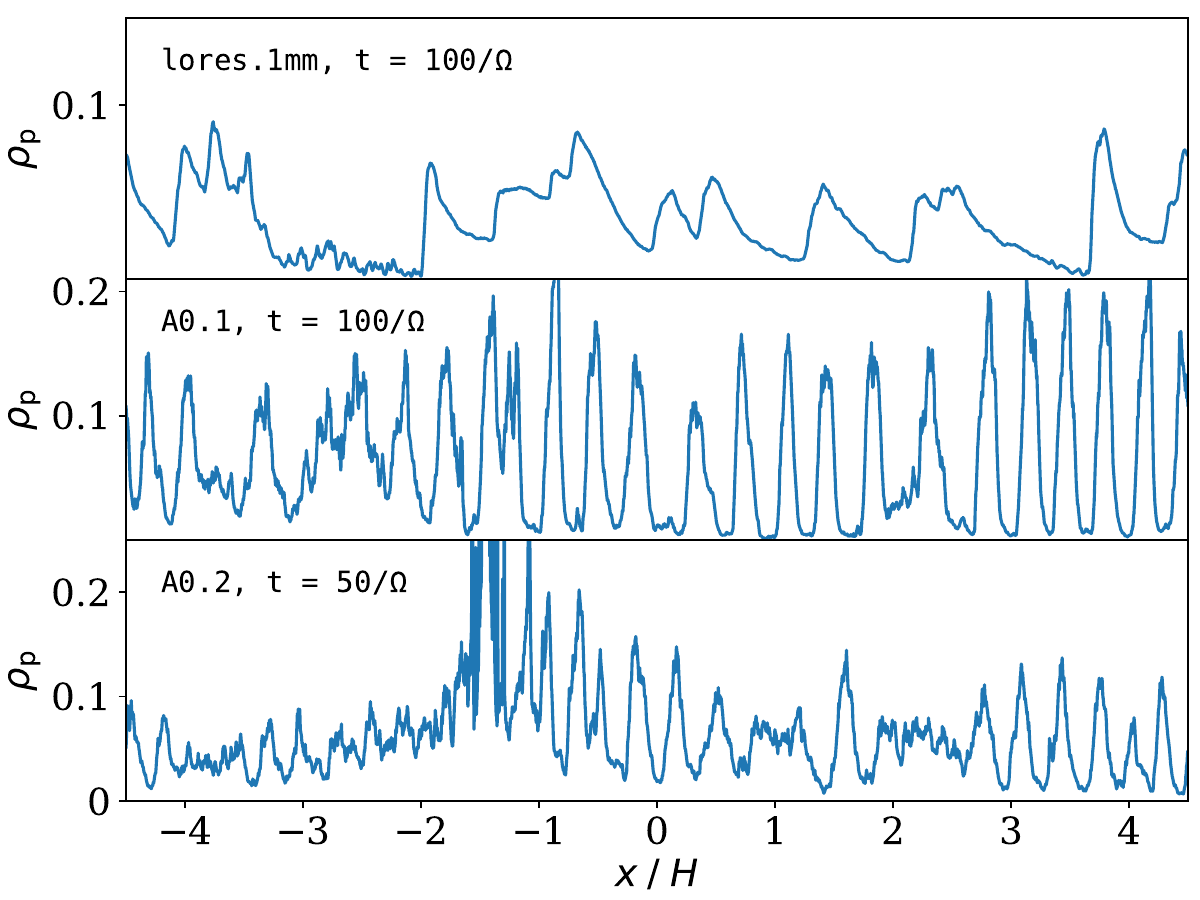}
    \caption{Mean particle density across the shearing box for three sample simulations in code units, where the gas density is $\rho \approx 1$. Even simulations that do not form planetesimals --- top two plots --- consistently form particle filaments. Furthermore, particle filaments are not restricted to the vicinity of the pressure bump ($|x/H| \lesssim 2$). Rather, filaments form everywhere and as they pass through the bump the filament density increases (e.g. bottom plot).}
    \label{fig:filaments}
\end{figure}

One unexpected result is that all of our simulations --- including those that did not form planetesimals --- consistently formed particle filaments. Moreover, the filaments form everywhere and are not restricted to the vicinity of the pressure bump. Figure \ref{fig:compare_A01} shows a 2D view of some of the filaments in \texttt{A0.1}. Figure \ref{fig:filaments} shows the dust density profile for snapshots of two runs that did not form planetesimals (\texttt{lores.1mm} and \texttt{A0.1}), along with our fiducial run (\texttt{A0.2}). This is important because the traditional clumping criteria of \citet{Carrera15,Yang_2017} would predict that these regions of the box \textit{should not be able to form particle clumps}. The dust-to-gas ratio at the edges ($Z \sim 0.01$) is far too small for the SI to be efficient --- especially for \texttt{lores.1mm} ($\tau \sim 10^{-2}$). Yet, filaments form. Evidently, clumping criteria obtained from small shearing-box simulations may not generalize to large slices of the disk.

We find that the pressure bump does not \textit{cause} the formation of SI filaments, but rather, it causes filaments to become \textit{denser} when they pass through the bump. This effect is hard to see on a still image, but it's most visible in the bottom plot (\texttt{A0.2}) of Figure \ref{fig:filaments} --- note that the filaments in the planetesimal formation region ($-2 \lesssim x/H \lesssim 0$) have a much higher peak than those near the edges of the box.

\subsection{Feedback and Reinforcement}
\label{sec:results:reinforcement}

The same particle feedback that is responsible for the SI also disrupts the pressure bump. As particles push back on the gas, they dissipate the pressure bump and alter its shape and location. So far in this work we have invoked an unspecified external force (such as a planet) to regularly reinforce the pressure bump (section \S\ref{sec:setup:T_reinf}). In this section we explore how the pressure bump responds to particle feedback in the absence of external reinforcement.

\begin{figure*}
    \centering
    \includegraphics[width=0.47\textwidth]{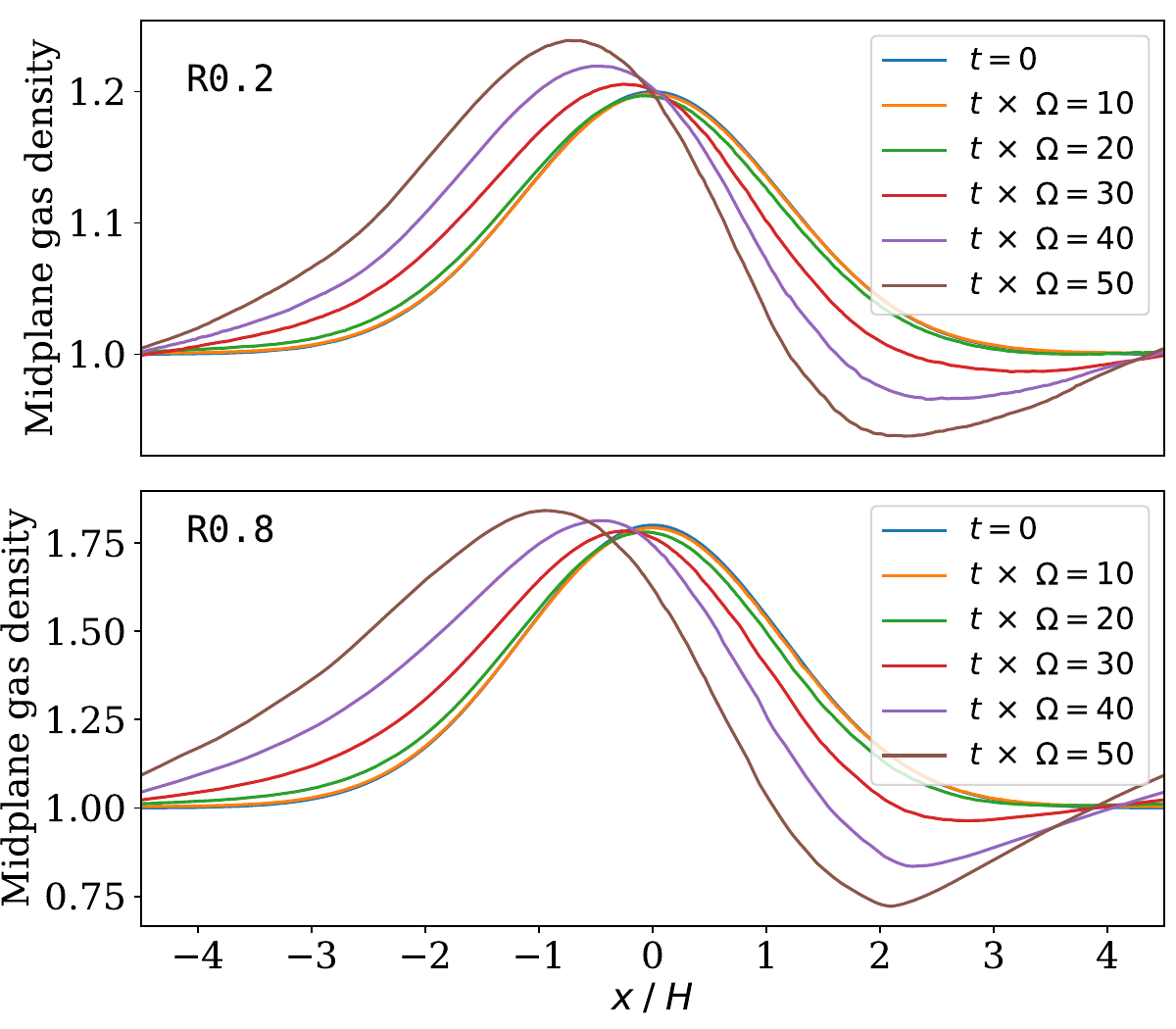}
    \includegraphics[width=0.47\textwidth]{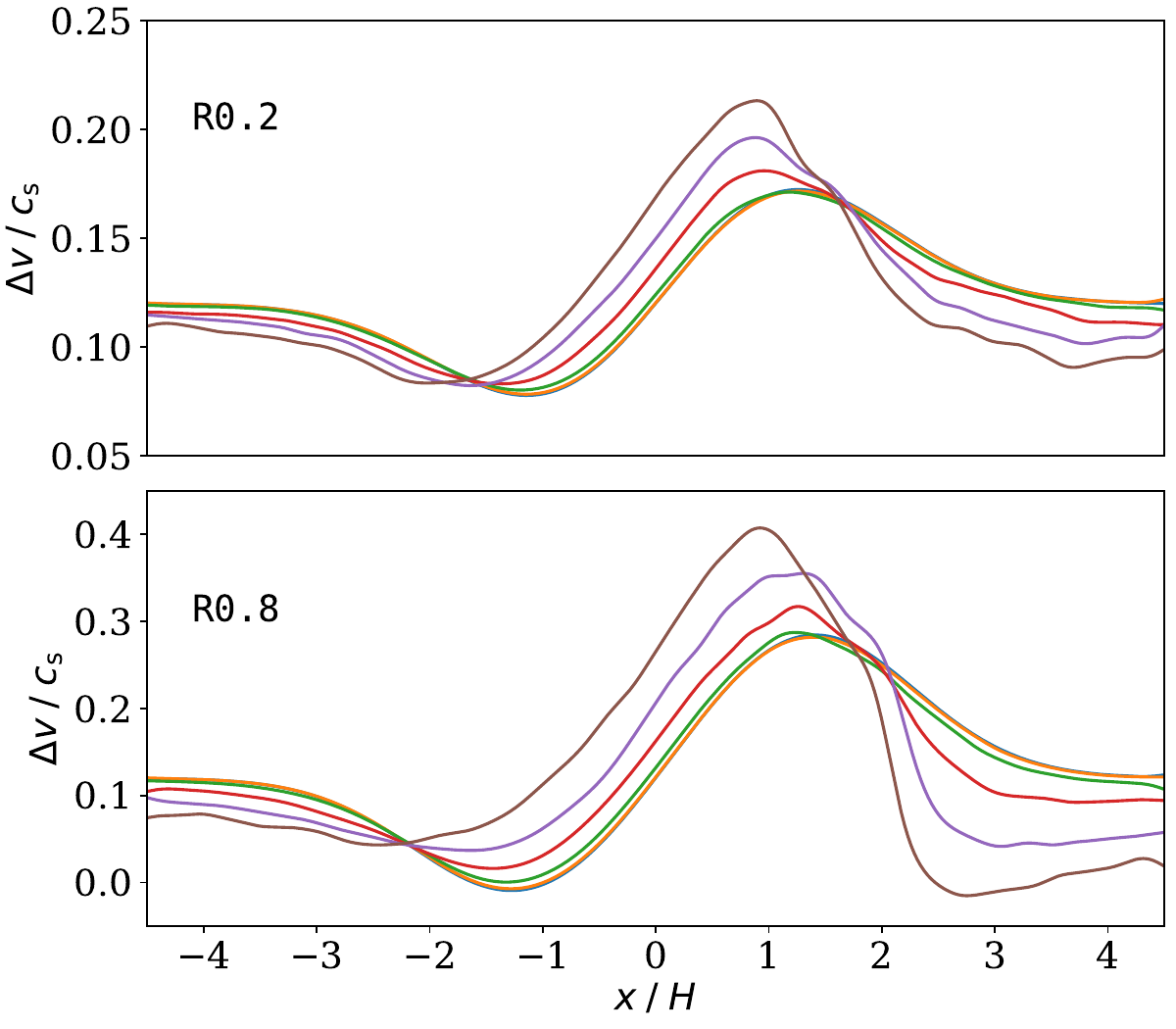}
    \caption{\textit{Left:} Snapshots of the mid-plane gas density profile (in code units) for two simulations with no pressure bump reinforcement. \textit{Right:} Azimuthally averaged headwind for the same snapshots. The runs begin with a fully formed pressure bump with amplitude $A = 0.2$ (top) or 0.8 (bottom) in geostrophic balance. In the absence of particle feedback, the pressure bump would survive indefinitely. With particles present, particle feedback gradually dissipates the bump, skews the bump toward the star, pushes the peak inward, and distorts the gas velocity profile. This occurs over $\sim 50 \Omegainv$, and many features are largely independent of the bump size.}
    \label{fig:bump_evo}
\end{figure*}

Simulations \texttt{R0.2} and \texttt{R0.8} begin with a fully formed pressure bump with amplitudes of $A = 0.2$ and 0.8. At the beginning of each simulation the pressure bump is in geostrophic balance, meaning that in the absence of particle feedback the bump would be sustained indefinitely. Figure \ref{fig:bump_evo} shows the midplane gas density for \texttt{R0.2} and \texttt{R0.8}. Over the first $\Delta t \sim 10 \; \Omegainv$ there is very little change because the particle density (and thus, particle feedback) is initially low. For the first $t \sim 10 \; \Omegainv$ the particles sediment (Figure \ref{fig:d_max}). At that moment, the particle density at the midplane is high enough to alter the azimuthal velocity of the gas and, without reinforcement, the shape of the bump begins to change.

Figure \ref{fig:bump_evo} shows how the pressure bump is gradually dissipated by particle feedback, and the shape of the bump visibly changes over the scale of $\sim 10 \; \Omegainv$. While that happens, the bump also drifts inward and gains a negative skew. We were surprised that runs \texttt{R0.2} and \texttt{R0.8} have very similar bump evolution, despite the bumps having very different sizes. Our interpretation is that, while a small pressure bump may, in principle, be more vulnerable to disruption by particle feedback, the amount of feedback is proportional to the particle concentration, which itself is driven by the bump size. In other words, it appears that pressure bump dissipation by particle feedback is a somewhat self-similar process.

The negative skew in figure \ref{fig:bump_evo} is important because the star-ward side of the bump (i.e., interior to the peak) is also where particles would normally concentrate to form planetesimals. We find that in \texttt{R0.2} and \texttt{R0.8} planetesimal formation is less efficient than in \texttt{A0.2} and \texttt{A0.8} respectively. Inspired by \citet{Lenz_2019}, we define the planetesimal formation efficiency $\epsilon$ by

\begin{equation}\label{eqn:epsilon}
    \epsilon \approx f \cdot \frac{L_x / T}{\vr}
\end{equation}

\noindent
where $f$ is the fraction of the particle mass at the end of the simulation that will very likely be converted into planetesimals (we chose particles in grid cells with $\rhop > 10^3 \rho_{\rm roche}$), $L_x$ is the length of the simulation box, $T$ is the simulation time, and $\vr$ is the \textit{unperturbed} radial drift speed. In other words, the effect the pressure bump has on $\vr$ is included in $\epsilon$.\footnote{We did one test where we defined $\vr$ as the average particle drift rate across the entire simulation (\texttt{A0.2}) and the value was nearly identical.}

The intuitive explanation for Equation \ref{eqn:epsilon} is that, if $\vr = L_x / T$ that means that, in principle, the simulation has run long enough to allow every particle a chance to pass through the pressure bump and potentially become a planetesimal, and in that case $f$ is a good estimate of the final planetesimal formation efficiency. In practice, none of our simulations run this long, so we attempt to extrapolate. In other words, $\epsilon$ is a simple extrapolation of $f$; it is a rough estimate of the planetesimal formation efficiency in a disk with bumps separated by distance $L_x$ where every particle gets a chance to pass through the pressure bump once.

One salient limitation of Equation \ref{eqn:epsilon} is that this kind of extrapolation is invalid in runs where particles can drift outward. Therefore, we do not compute $\epsilon$ for runs \texttt{A0.8} and \texttt{R0.8}. Table \ref{tab:efficiency} shows $\epsilon$ and $f$ for all the runs that produced bound clumps.

\begin{table}[h!]
\centering
\caption{Planetesimal formation efficiency $\epsilon$ (Equation \ref{eqn:epsilon}) and fraction of the particle mass in bound clumps $f$ for every model that made bound clumps. Lack of reinforcement in \texttt{R0.2} and \texttt{R0.8} substantially reduces the formation efficiency; especially at low amplitude (\texttt{R0.2}). There is no $\epsilon$ for \texttt{A0.8} and \texttt{R0.8} because Equation \ref{eqn:epsilon} is not applicable to a model where particles can drift outward.}
\begin{tabular}{ccc}
Model & Planetesimal formation efficiency $\epsilon$ & $f$\\
\texttt{A0.2} & 0.42 & 0.08 \\
\texttt{A0.4} & 0.87 & 0.10 \\
\texttt{A0.8} & --- & 0.15 \\
\hline
\texttt{R0.2} & $9.6 \times 10^{-3}$ & $3.2 \times 10^{-3}$ \\
\texttt{R0.8} & --- & 0.11
\end{tabular}
\label{tab:efficiency}
\end{table}

\begin{figure}[ht!]
    \centering
    \includegraphics[width=0.47\textwidth]{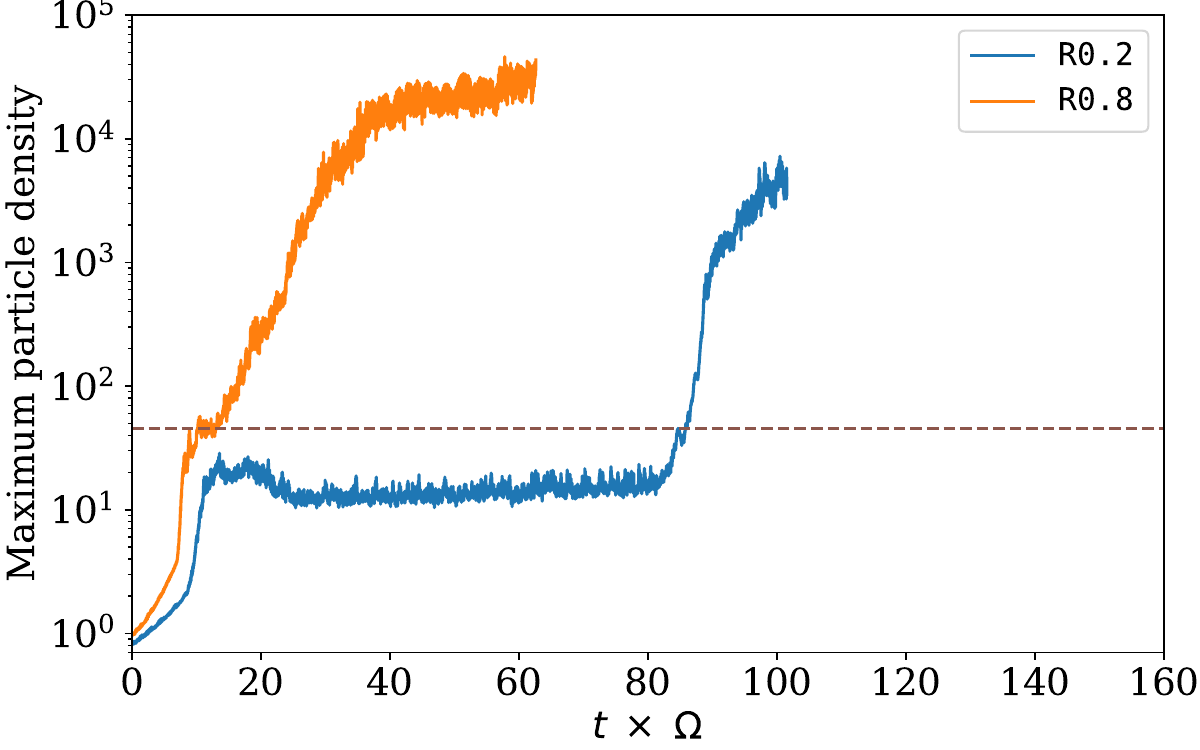}
    \caption{Maximum particle density (in code units) versus time for the two runs with no bump reinforcement. The dashed line marks the Roche density. Despite the lack of reinforcement, both runs reach the Roche density and form gravitationally bound clumps, though with a notable delay in the case of \texttt{R0.2} (compare with figure \ref{fig:d_max}).}
    \label{fig:d_max_noreinf}
\end{figure}

Figure \ref{fig:d_max_noreinf} shows the maximum particle density for \texttt{R0.2} and \texttt{R0.8}. Run \texttt{R0.2} forms planetesimals much later than \texttt{A0.2}. Run \texttt{R0.8} crosses the Roche density sooner than \texttt{A0.8}, partly aided by a bump that is already formed, but the density growth is slower and the peak density is reached later than in \texttt{A0.8}.

The relevance of the comparison with \texttt{R0.2} and \texttt{R0.8} is limited because it seems unlikely that a pressure bump would form, but then have no mechanism to reinforce it. A more physically realistic scenario would be to model a bump that develops with a reinforcement timescale longer than $\treinf = 1\;\Omegainv$. We will explore that idea in a future investigation.

Our finding that particle feedback significantly alters the pressure bump is in conflict with \citet{Onishi17}. They found that particles sediment to a thin layer so that most of the gas in the bump is unaffected by the back-reaction. The discrepancy cannot be due to the box size since our box is in fact taller ($z_{\rm max} = 0.2 H$ vs $0.125 H$) and our particles are larger ($\tau \approx 0.12$ vs $0.01$) than theirs. Other differences between \texttt{R0.2} and \citet{Onishi17} include the background pressure gradient $\Pi$, the presence of an azimuthal direction, and boundary conditions --- they use a reflecting boundary in the vertical direction; we use an open boundary with regular density rescaling to conserve mass. It is possible that some of these might be responsible for the different results.

\subsection{Characteristic Particle Concentration}
\label{sec:results:dpar_histogram}

\begin{figure}
    \centering
    \includegraphics[width=0.47\textwidth]{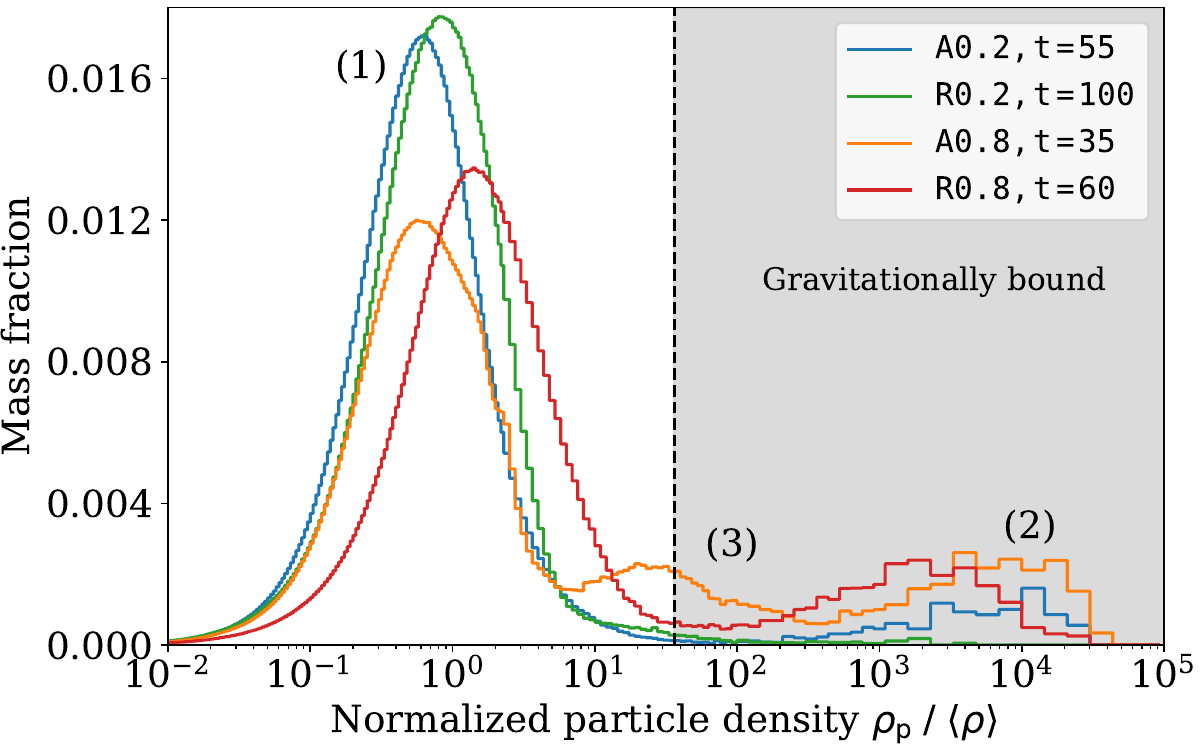}
    \caption{Histogram of the particle density near the end of four simulations. Two runs have a bump amplitude of $A = 0.2$ (\texttt{A0.2} and \texttt{R0.2}) and two have $A = 0.8$. In two runs, the pressure bump is reinforced on a timescale of $\treinf = 1\Omegainv$ (\texttt{A0.2} and \texttt{A0.8}) and in the other runs the bump is not reinforced at all (\texttt{R0.2} and \texttt{R0.8}). The main mode (1) shows that most of the particle mass is in regions with $\rhop \sim 1$ (i.e. comparable to the gas density) and is probably associated with streaming instability filaments. All runs have a second mode (2) at $\rhop \sim 10^4$, probably dominated by self-gravity. Run \texttt{A0.8} has another mode (3) at $\rhop \sim 40$, likely caused by a combination of the particle trap and the streaming instability.}
    \label{fig:dpar_histogram}
\end{figure}

Figure \ref{fig:dpar_histogram} shows a histogram of the particle density for runs \texttt{A0.2}, \texttt{A0.8}, \texttt{R0.2}, and \texttt{R0.8}. It shows the fraction of the particle mass in each density interval. Perhaps the most interesting feature of this graph is that it clearly shows at least two characteristic density scales that are present in all simulations:

\begin{description}
\item[(1)]  The bulk of the particle mass is always inside a dominant mode in the vicinity of $\rhop \sim 1$, where the particle density is in the same magnitude range as the gas density. This is likely the characteristic density scale of streaming instability filaments.

\item[(2)]  There is always a second mode in the vicinity of $\rhop \sim 10^3-10^4 \gg \rho_{\rm roche}$, where the physics is clearly dominated by self-gravity. The runs with  bump reinforcement (\texttt{A0.2}, \texttt{A0.8}) have higher $\rhop$ than their un-reinforced counterparts. We shall return to this point at the end of this section. Note that that mode is still present for \texttt{R0.2} though it is somewhat difficult to see in the figure as it contains only 1\% of the particle mass.
\end{description}

\begin{figure}
    \centering
    \includegraphics[width=0.47\textwidth]{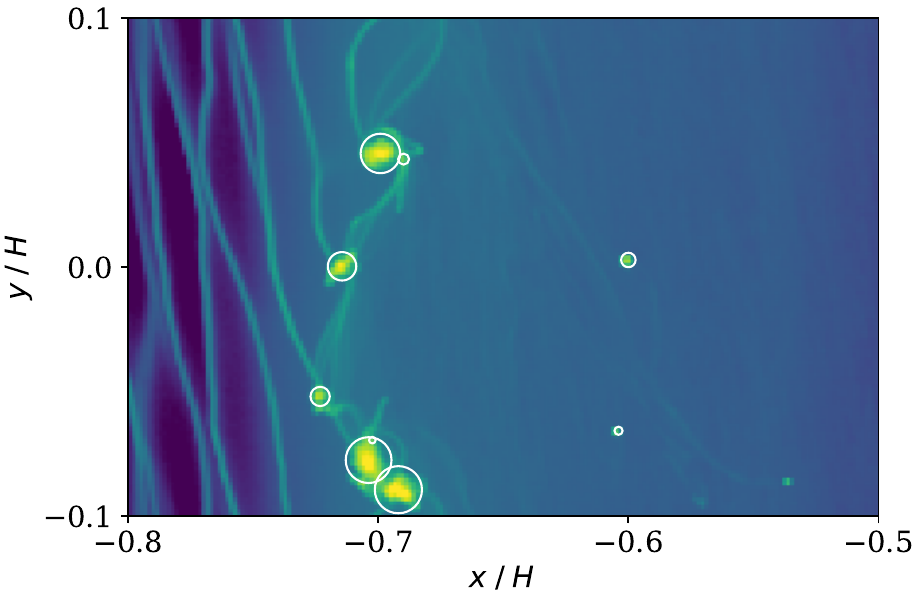}
    \caption{Column  dust/gas ratio ($Z = \Sigmap / \Sigma$) for a narrow slice of model {\tt A0.8} near the end of the simulation. Unique to this run is another prominent structure --- a network of thin, dense filaments, often connecting bound particle clumps to one another. This structure is associated with the third mode \textbf{(3)} in figure \ref{fig:dpar_histogram}.  The color bar employed is the same as the previous snapshots (Figs.~\ref{fig:formation_A02}-\ref{fig:compare_A02_NFB}). } 
    \label{fig:zoom_A08}
\end{figure}

Finally, run \texttt{A0.8} has a third major mode \textbf{(3)} near $\rhop \sim 40$ that does not seem to be present in the other runs. This points to an additional structure that is likely associated with the particle trap (\texttt{R0.8} begins with a particle trap too, but it moves as the bump is deformed). Figure \ref{fig:zoom_A08} shows a close-up view of the planetesimal forming region of run \texttt{A0.8}. There is indeed an additional layer of structure that we don't generally see in other runs: A network of very thin and dense particle filaments, often connecting particle clumps to one another. Similar filaments can be seen in \texttt{A0.4} at $t = 30/\Omega$ (Figure \ref{fig:formation_A04}) but they appear to be a transient feature that is quickly destroyed by Keplerian shear. This filamentary structure can only last if the bound clumps form in nearly the same orbit, as is the case in \texttt{A0.8}.

The bound clumps in runs with bump reinforcement are more massive than their un-reinforced counterparts. This likely explains the higher characteristic density in \texttt{A0*} vs \texttt{R0*} (Figure \ref{fig:dpar_histogram}). Quantitatively, the mean clump mass in \texttt{A0.8} (1.58\% of the total particle mass) is 22 times larger than that of \texttt{R0.8} (0.07\%) and the mean clump mass in \texttt{A0.2} (0.41\%) is 9 times larger than that of \texttt{R0.2}. In addition, \texttt{A0.8} stands out because it is dominated by a handful of very massive clumps --- it has four clumps with $3.4\%$ of the total particle mass and five smaller clumps with $0.1\%$ of the particle mass. Run \texttt{A0.2} forms 19 clumps with a more equal distribution of mass. It is possible that the presence of filaments (as discussed above) and the slower radial drift (see Fig.~\ref{fig:formation_A08}) in \texttt{A0.8} conspire to continually feed the clumps already formed, leading to several very massive clumps.

\section{Uncertainties and Future Work}
\label{sec:uncertainties}

Our work is subject to a number of uncertainties and limitations, both numerical and physical in nature.  First, as with many previous studies of streaming-induced planetesimal formation, our FFT-based gravity solver halts collapse at the grid scale, which prevents our planetesimals to collapsing to scales $\sim 100$~km. While this limitation restricts what we can learn about the physical properties of planetesimals, it likely does not play a significant role (if any at all) in whether or not planetesimals form as well as how and when they form, which are the primary questions addressed in this work.  

Furthermore, as pointed out in Section~\ref{sec:setup:resolution}, the radial extent of our domain stretches the validity of the local approximation.  However, as we also discussed above, the particle enhancement and subsequent growth of the streaming instability occurs on scales $\ll L_x$.  In the absence of global simulations that incorporate the same physics (which are not yet developed), such extended local simulations will suffice. 

An additional limitation associated with this setup is the approach used to simulate a pressure bump: the pressure bump is artificially induced and maintained via Newtonian relaxation towards a state of geostrophic balance. While this configuration gives us explicit control over the relevant bump parameters (e.g., amplitude, reinforcement timescale), and in that sense can be seen as a strength, future simulations that include bump-inducing mechanisms (e.g., planets, magnetically-induced zonal flows) will be necessary in order to fully test the results explorable with our current setup. 

Additionally, most of our simulations assume cm-size particles. In regions where drift-limited particle growth occurs (e.g., \citealt{Birnstiel12}), these particles may be too large. However, the simulations carried out here provide invaluable insight even in such a limit, and our tentative result that mm-sized particles do not produce strong clumping strongly motivates future studies to determine whether or not this is indeed the case.

Finally, in this same vein, a recent result by \cite{Krapp19} shows that the linear growth of the SI may not converge with increasing number of particle sizes.  If such non-convergence carries over to the nonlinear regime in the absence of a pressure bump, studies {\it with} a pressure bump may be crucial as such bumps serve as natural sites to spatially separate particles of different sizes (e.g., smaller $\tau$ particles will be less concentrated within the bump in the presence of turbulent diffusion).

\section{Summary and Conclusions}
\label{sec:conclusions}

We conducted shearing box simulations with {\sc Athena} in order to explore the formation of planetesimals in a pressure bump similar in size to those observed by ALMA in nearby protoplanetary disks \citep[e.g.][]{Huang18}. Previous numerical work on the streaming instability has relied on simulations with idealized conditions, such as a large initial solid-to-gas ratio $Z$, that are ``rigged'' to be more conducive to planetesimal formation. Our work presents the first set of simulations that show that planetesimal formation is possible (and even quite robust) under conditions likely to be realized in protoplanetary disk systems, namely $Z = 0.01$ (comparable to the solar nebula) and in the presence of largely axisymmetric (but RWI-stable) pressure bumps that have properties similar to those observed by ALMA. Our main conclusions are as follows.

\begin{enumerate}
\item Planetesimal formation inside a pressure bump is a robust process that does not require a particle trap and can occur even for moderately low amplitude pressure bumps that are more likely to be RWI-stable. The local enhancement in particle density that arises from particles drifting through the pressure bump is sufficient to kick-start the streaming instability and produce planetesimals.

\item As a corollary, planetesimal formation occurs in filaments that are continually drifting inward past the pressure bump maximum; planetesimals are formed inward (i.e., closer to the star) of the pressure bump that initiated their growth.

\item These planetesimals are formed via the streaming instability and not direct gravitational collapse due to concentration from the pressure bump itself.

\item Particle filaments are a much more robust process than traditional clumping criteria suggest \citep{Carrera15,Yang_2017}. Filaments form across the entire box, even in simulations that do not form planetesimals and even in regions where the clumping criterion is not satisfied. Evidently, the clumping criterion from small box simulations does not seem to generalize to larger scales.

\item Bound particle clumps from simulations with a particle trap (run \texttt{A0.8}) are fewer in number, less radially dispersed, and more massive. 

\item In the absence of reinforcement, feedback from the particles distorts the shape of the pressure bump. Nonetheless, planetesimals still form, albeit at a lower efficiency.

\item The critical bump amplitude needed to trigger planetesimal formation (for the parameters combinations considered here) appears to be higher than 10\% and below 20\%. This value may be different in the inner disk, where the background pressure gradient is less steep, and it may also be different for small particles.

\item For the resolution employed here, mm sized particles do not produce planetesimals in pressure bumps. Given that at large radial distances from the star, drift limited particle growth limits particle sizes to mm (see \citealt{Birnstiel12}), this result, if it holds at higher resolutions, suggests that planetesimal formation is not possible within the drift limited regions of protoplanetary disks.
\end{enumerate}

Taken together, these results have one underlying implication: planetesimal formation in pressure bumps from cm-sized particles occurs via the SI and is extremely robust.  While future work to study different bump reinforcement timescales will be required to verify this result, the fact that amplitudes of only $\sim 10$--20\% are needed to initiate the streaming instability points to a wide variety of conditions under which planetesimals can form.  

The most studied mechanism for forming pressure bumps is the carving of gaps by planets in protoplanetary disks.  As the planet exchanges angular momentum with the gas, a gap opens up, which ultimately leads to a pressure increase (a ``bump") moving away from the gap. Indeed, there is now observational evidence that the gaps observed in many Class II sources (e.g., \citealt{Huang18}) are actually the result of planets \citep{Pinte18,Teague19}. Of course, if planetesimals form in these planet-induced bumps, they would represent a {\it later} generation of planetesimals, and not those that formed the first generation of planets.  

Saving us from the pitfalls of this ``chicken or the egg'' type paradox (i.e., how did the first generation of planetesimals form?), there is both theoretical and observational evidence that pressure bumps can form in even younger systems and without the aid of already-formed planets.  In terms of observations, both HL Tau (likely transitioning from Class I to Class II; \citealt{Alma15}) and GY 91 (a Class I system; \citealt{Sheehan18}) contain dust rings that indicate the presence of pressure bumps. While such observations do not rule out the planet hypothesis for bump formation, it motivates the consideration of alternative mechanisms, lest planet formation in these very young systems is far enough along that they are already able to carve substantial gaps. 

On the theoretical side, a large number of alternative mechanisms have been proposed to generate bumps and rings in disks.  For example, many studies that have explored the evolution of magnetically-induced zonal flows and related phenomena \cite[e.g.,][]{Lyra08,Johansen09a,Dzyurkevich10,Uribe11,Simon12,Simon14,Bai15,Suriano18,Riols19}. While the exact amplitude of the pressure bumps induced by these processes depends on the particular setup (e.g., assumed magnetic field strength),  it often exceeds $\sim 20\%$ \citep{Dzyurkevich10,Uribe11,Bai15}. 

Furthermore, transitions between ionization regions can also generate pressure bumps.  \citet{Dzyurkevich10} showed that the inner edge of the Ohmic dead zone (see \citealt{Gammie96} for a description of the dead zone) may harbor a pressure bump much larger than 20\%.  \cite{Flock15} carried out similar calculations and found that a pressure bump can form at the outer edge of the dead zone with sufficiently large amplitude as to halt radial drift of particles (though, consistent with our discussion above, this bump was unstable to the RWI). 

Beyond these models, there are a number of other mechanisms that can produce pressure bumps in disks, many of which have not been characterized in terms of percentage of pressure variation.  However, amplitudes of $\sim 10$--20\% are not outrageously high; indeed, one can easily envision any number of mechanisms producing such modest amplitude fluctuations.  Ultimately, the prevalence of ring structures, both in observational detections and mechanisms by which they can form, coupled with the results of this work strongly indicate that planetesimal formation in pressure bumps is not only viable, but is very likely commonplace.

\added{\software{{\sc Athena} \citep{Stone08,Bai10a,Simon16a}, Julia \citep{Julia-2017}, 
and Matplotlib \citep{Matplotlib}.}}

\acknowledgements
We thank Hui Li, Jonathan Squire, Phil Armitage, Andrew Youdin, and Wlad Lyra for useful discussions and suggestions regarding this work. We also thank the referee for all the useful feedback that helped improve this manuscript. DC and JBS acknowledge support from NASA under {\em Emerging Worlds} through grants 80NSSC18K0597 and 80NSSC19K0502. RL acknowledges support from NASA under grant NNX16AP53H. The numerical simulations and analyses were performed on {\sc Stampede 2} through XSEDE grant TG-AST120062.

\bibliographystyle{apj}
\bibliography{references}

\end{document}